\documentclass[a4paper,11pt]{article}
\usepackage{jheppub}
\usepackage{bm, comment, braket}
\usepackage[x11names]{xcolor}
\usepackage[T1]{fontenc}
\usepackage{colortbl}
\newcommand{\del}{\partial}
\newcommand{\beq}{\begin{eqnarray}}
\newcommand{\eeq}{\end{eqnarray}}

\usepackage{amsmath}
\allowdisplaybreaks
\usepackage{ulem}
\usepackage{cancel}

\newcommand{\pd}{\partial}
\newcommand{\mkk}{\mathcal{M}_{\mathrm{KK}}}
\newcommand{\ukk}{u_{\mathrm{KK}}}
\DeclareMathOperator{\tr}{tr}

\DeclareMathOperator{\am}{am}
\DeclareMathOperator{\dn}{dn}

\begin{document}

\title{
Chiral Soliton Lattices under Magnetic Fields and Rotation: a Holographic Analysis
}

\author[a]{Markus~A.~G.~Amano,}
\emailAdd{markus(at)oyama-ct.ac.jp}

\author[b,c,d]{Minoru~Eto,}
\emailAdd{meto(at)sci.kj.yamagata-u.ac.jp}

\author[e,c,d]{Muneto~Nitta,}
\emailAdd{mune.nitta(at)gmail.com}

\author[f]{
and Shin~Sasaki}
\emailAdd{shin-s(at)kitasato-u.ac.jp}

\affiliation[a]{
Department of Innovative Electrical \& Electronic Engineering, National Institute of Technology, Oyama College, Ooazanakakuki 777, Oyama City, Tochigi, 323-0806, Japan
}

\affiliation[b]{
Department of Physics, Yamagata University, Kojirakawa-machi 1-4-12, Yamagata, Yamagata
990-8560, Japan
}

\affiliation[c]{
Research and Education Center for Natural Sciences,
Keio University, 4-1-1 Hiyoshi, Kanagawa 223-8521, Japan
}

\affiliation[d]{
International Institute for Sustainability with Knotted Chiral Meta
Matter(WPI-SKCM$^2$), Hiroshima University, 1-3-2 Kagamiyama,
Higashi-Hiroshima, Hiroshima 739-8511, Japan
}

\affiliation[e]{
Department of Physics, Keio University, 4-1-1 Hiyoshi, Yokohama,
Kanagawa 223-8521, Japan
}

\affiliation[f]{
Department of Physics, Kitasato University, Sagamihara 252-0373, Japan
}

\abstract{
We study the chiral soliton lattice (CSL) in rotating QCD matter under a background magnetic field within holographic QCD.
We show that rotation can be incorporated as a background gauge field and that the CSL in rotating matter becomes a ground state in the gravity dual.
We also provide a brane interpretation of the CSL under a magnetic field and rotation.
Solving the bulk equations, we derive an effective Hamiltonian where the meson decay constant 
$\tilde{f}$ emerges as an anisotropic, field-dependent matrix.
Furthermore, we discuss the thermodynamic properties of the ground state in rotating matter and study its equation of state.
}

\maketitle

\section{Introduction}

Quantum Chromodynamics (QCD) under extreme conditions provides a rich
arena for nonperturbative phenomena.
In particular, dense QCD matter exposed to strong magnetic fields and/or
rapid rotation is relevant to strongly magnetized compact stars and
relativistic heavy-ion collisions \cite{Fukushima:2010bq,Schmidt:2017bjt}.
At low energies, the dynamics of the Nambu--Goldstone modes associated
with spontaneous chiral symmetry breaking is described by chiral
perturbation theory (ChPT) \cite{Scherer:2012xha, Bogner:2009bt}.
An essential ingredient in external backgrounds is the
Wess--Zumino--Witten (WZW) term
\cite{Wess:1971yu,Witten:1983tx,Harvey:2007ca}, which encodes the quantum anomalies of
the microscopic theory.
At finite density, the gauged WZW term generates anomalous couplings
between neutral mesons and electromagnetic, baryon-number, and isospin
background gauge fields \cite{Son:2004tq,Amano:2026jsv}.
These couplings underlie anomalous charges and currents carried by
topological defects \cite{Goldstone:1981kk,Fukushima:2018ohd} and,
more importantly for the present work, can alter the ground state itself
by energetically favoring spatially modulated configurations
\cite{Son:2007ny}.

A paradigmatic example is the chiral soliton lattice (CSL) in dense QCD
under a magnetic field.
For sufficiently large baryon chemical potential and magnetic field,
the anomaly-induced term linear in the spatial derivative of the neutral
pion favors a periodic array of sine-Gordon solitons carrying baryon-number
density \cite{Son:2007ny,Eto:2012qd,Brauner:2016pko}.
The properties of this phase have been investigated from several
complementary viewpoints, including thermal fluctuations and
finite-temperature effects
\cite{Brauner:2017uiu,Brauner:2017mui,Brauner:2021sci,
Brauner:2023ort}, quantum, thermal, and dynamical nucleation
\cite{Eto:2022lhu,Higaki:2022gnw,Eto:2025ebz}, and coupled
pion--$\eta$ configurations and incommensurate crystals
\cite{Qiu:2023guy}.
Closely related CSL phases have also been studied in QCD-like theories
without a sign problem
\cite{Brauner:2019rjg,Brauner:2019aid} and in supersymmetric QCD
\cite{Nitta:2024xcu}.

The anomaly-induced phase structure is not limited to one-dimensional
pion modulations. 
Larger chemical potential and/or magnetic fields stabilize domain-wall Skyrmions, in which baryonic
solitons reside on the constituent walls of a CSL
\cite{Eto:2023lyo,Eto:2023wul,Amari:2024fbo,Amari:2024mip,Copinger:2025rpo}.
For a relation between domain-wall Skyrmions and bulk Skyrmions, see  refs.~\cite{Kawaguchi:2018fpi,Chen:2021vou,Chen:2023jbq, Amari:2025twm}.
Other proposed inhomogeneous states include baryonic crystals
\cite{Evans:2022hwr,Evans:2023hms} and vortex-Skyrmion or
baryonic-vortex phases at nonzero isospin density
\cite{Gronli:2022cri,Qiu:2024zpg,Hamada:2025inf,Hamada:2026uec,
Mameda:2026kyp}.
See also Refs.~\cite{Canfora:2018rdz,
Canfora:2020uwf,
Canfora:2020kyj,
Canfora:2024mkp} for related works.

A common feature of these phases is that anomaly-induced topological terms
endow their solitonic constituents with baryon number and can make them
energetically competitive with homogeneous matter.

Rotation gives rise to a parallel, but distinct, class of anomaly-driven
phenomena.
Rapidly rotating dense QCD can support CSLs built from the $\eta$ or
$\eta'$ sector \cite{Huang:2017pqe,Nishimura:2020odq}, as well as
non-Abelian CSLs, domain-wall Skyrmion phases, and related low-energy
excitations
\cite{Chen:2021aiq,Eto:2021gyy,Eto:2023tuu,Eto:2023rzd,
Evans:2025jsa}.
At finite chemical potential, rotation may be represented, to leading
order in the local velocity, by spatial components of a background
$\mathrm{U}(1)$ gauge field whose effective magnetic field is proportional to the
chemical potential times the vorticity
\cite{Son:2004tq,Abramchuk:2018jhd}.
This identification is closely related to the Barnett effect in
relativistic and nuclear matter
\cite{Becattini:2020nfu,Becattini:2021lfq,Huang:2017pqe}, and
analogous local-boost constructions have been employed in holographic
settings \cite{Zhao:2022uxc,Chen:2020ath}.
It therefore provides a convenient way of treating magnetic and vortical
anomaly-induced effects on a common footing.

To investigate these phases beyond the regime of the derivative
expansion, we employ holographic QCD, in particular the
Sakai--Sugimoto model \cite{Sakai:2004cn,Sakai:2005yt}.
The model gives a gravitational description of large-$N_c$ QCD in which
spontaneous chiral symmetry breaking is realized geometrically by the
joining of flavor D8- and $\overline{\mathrm{D8}}$-branes, while the
anomalous WZW couplings descend from the five-dimensional Chern--Simons
term on their worldvolume.
Holographic QCD at finite baryon density and in external magnetic fields
has been explored in a variety of contexts
\cite{Rebhan:2008ur,Preis:2011sp,Evans:2022hwr,Callebaut:2011ab,
Burikham:2011rg,Fukushima:2013zga,Bartolini:2023wis,
Bartolini:2023eam,Kovensky:2023mye};
see also Refs.~\cite{Thompson:2008qw,Bergman:2008qv} for related
analyses of magnetic backgrounds in the Sakai--Sugimoto model.

In our previous holographic study \cite{Amano:2025iwi}, we demonstrated
that the neutral-pion CSL in a background electromagnetic magnetic field
is realized as a ground state of the gravity dual once appropriate
boundary conditions and a quark-mass deformation are imposed.
We further showed that the CSL admits a brane interpretation in terms of
uniformly distributed D4-brane charge and five-dimensional instanton
density, and that the bulk dynamics renders the effective pion decay
constant magnetic-field dependent.
In a complementary recent analysis \cite{Amano:2026jsv}, we systematically
revisited the anomaly-induced WZW couplings of neutral mesons in nuclear
and quark matter under magnetic fields and rotation.
The resulting terms proportional to
$\vec B\cdot\vec\nabla\phi$ and
$\vec\Omega\cdot\vec\nabla\phi$ ($\phi$ is the meson)
provide the low-energy starting point for studying CSLs in simultaneous
magnetic and vortical backgrounds.

In this paper, we extend the holographic construction of
ref.~\cite{Amano:2025iwi} to dense QCD matter subject simultaneously
to a background electromagnetic magnetic field and rotation.
Our central observation is that, at leading order in the local velocity,
the rotational effect can be encoded by a background $\mathrm{U}(1)_B$ gauge
field with an effective magnetic component proportional to
$\mu_B\vec\Omega$.
This makes it possible to incorporate the anomaly-induced effects of
magnetic fields and rotation within a unified five-dimensional
holographic description.

For $N_f=1$, we show that the $\eta$ field obeys a chiral sine-Gordon
model and forms a rotational CSL above a critical value of a combination
of the baryon chemical potential and angular velocity.
We identify the corresponding Ramond--Ramond charges and show that,
in contrast to the purely electromagnetic case, the rotational
configuration carries both dissolved D6-brane and D4-brane charges.
For $N_f=2$, we analyze the coupled neutral-pion and $\eta$ sectors in
simultaneous magnetic and rotational backgrounds.
The system separates into two chiral sine-Gordon sectors, leading to the
QCD vacuum, two single-CSL phases, and a dual-CSL phase.
We also clarify their brane interpretation in terms of D6-brane charges
and five-dimensional instanton density, the latter representing the
baryon-number density of the CSL.

We then solve the bulk gauge-field equations in the presence of general
magnetic and vortical backgrounds, thereby including the effects of the
tower of massive vector mesons.
The resulting on-shell Hamiltonian is characterized by an anisotropic,
background-dependent matrix of meson decay constants, generalizing the
magnetic-field-dependent decay constant found in
ref.~\cite{Amano:2025iwi}.
We determine the phase structure for massless and massive mesons and
compare the holographic phase boundaries with those obtained in ChPT.
We also find distinct weak- and strong-background regimes, 
and discuss the thermodynamic properties and equation of state
of the resulting ground states.

The organization of this paper is as follows.
In the next section, we briefly introduce the holographic setup of QCD
and its low-energy reduction to ChPT.
In sec.~\ref{sec:rotation}, we formulate rotation as a background
magnetic field of the $\mathrm{U}(1)_B$ baryon-number gauge field.
We first analyze the $N_f=1$ rotational CSL and its brane interpretation,
and then study the $N_f=2$ system in the simultaneous presence of an
electromagnetic magnetic field and rotation.
In sec.~\ref{sec:bulk_analysis}, we solve the bulk equations, derive the
effective Hamiltonian, and investigate the phase diagrams and
thermodynamic properties.
Section~\ref{sec:conclusion} is devoted to conclusions and discussion.
The reduction to ChPT in the massive case, the derivation of the
massless bulk solutions
are collected in the Appendices \ref{app:Reduction_to_ChPT} and \ref{app:massless_bulk_solutions}.

\section{Holographic setup} \label{sec:setup}
In this section, we briefly introduce the holographic QCD model 
based on $N_c$ color D4-branes and $N_f$ flavor D8-branes in Type IIA string theory 
-- the Sakai-Sugimoto model \cite{Sakai:2004cn,Sakai:2005yt}.
The model provides a gravity dual description of the large-$N_c$ QCD and reproduces the chiral perturbation theory (ChPT) of mesons on the worldvolume of the D8-branes.
Since our main interest lies in the phase structure of low-energy QCD matter under rotation and background magnetic fields, we focus on the effective action of the probe D8-branes.

\subsection{The model}
The model consists of the $N_c$ D4-branes and $N_f$ pairs of the D8/$\overline{\mathrm{D8}}$-branes ($N_f \ll N_c$) in ten-dimensional spacetime. 
The large number of color branes generates a geometry with a topology of $R^5\times S^1 \times S^4$ in the so-called confined phase.
These D4-branes extend along the directions $x^0$, $x^1$, $x^2$, and $x^3$, and are wrapped around a compactified circle direction $x^4$ with radius $\mkk^{-1}$. 
The anti-periodic boundary condition imposed on the fermions along the $x^4$ circle breaks supersymmetry, so one can neglect the fermionic degrees of freedom for energy scales less than $\mkk$.
Below the energy scale, $\mkk$, four-dimensional large $N_c$ $\mathrm{U}(N_c)$ massless QCD is realized on the D4-branes \cite{Sakai:2004cn,Sakai:2005yt}.

The background geometry generated by the $N_c$ D4-branes in the near-horizon limit is given by 
\begin{align}
 ds^2 =& \ 
 \left(
 \frac{u}{R}
 \right)^{3/2} 
 \Big(
 \eta_{\mu\nu} dx^\mu dx^\nu + f(u)d \tau^2
 \Big) 
 + 
 \left(
 \frac{R}{u}
 \right)^{3/2} 
 \left( 
 \frac{du^2}{f(u)} + u^2 d\Omega_4^2 
 \right), 
 \notag \\
 e^{\varphi} =& \ g_s \left(\frac{u}{R}\right)^{3/4}, 
 \quad 
 F^{(4)} = dC^{(3)} = \frac{2\pi N_c}{V_4} \epsilon_4, 
 \quad 
 f(u) = 1 - \frac{\ukk^3}{u^3},
 \label{eq:D4_geometry}
\end{align}
where $\eta_{\mu \nu} = \mathrm{diag} (-1,1,1,1)$ and 
$x^\mu (\mu=0,1,2,3)$ are the four-dimensional coordinates with flat metric, $x^4 \equiv \tau$ corresponds to the compactified direction with periodicity $\tau \sim \tau + 2\pi/\mkk$, and $u$ is the radial coordinate satisfying $u \ge \ukk$.
The fields $\varphi$, $C^{(p)}$, $F^{(p+1)}$ are the dilaton, the RR $p$-form and its field strength, respectively.
The quantities $d \Omega^2_4$, $\epsilon_4$, and $V_4$ represent the metric, the volume form, and the volume of $S^4$ surrounding the D4-branes.
The parameters $R$ and $\ukk$ are constants characterizing the geometry.
The Kaluza-Klein (KK) mass scale $\mkk$, which sets the scale of the theory, 
is related to the parameters by $\mkk = \frac{3}{2}\sqrt{\ukk/R^3}$. 
The 't Hooft coupling $\lambda$ is given by $\lambda = g_{\mathrm{DBI}}^2 N_c = 2 \pi \mkk g_s l_s N_c$ where $l_s^2 = \alpha'$ is the string length and $\alpha'$, $g_s$ are the string Regge parameter and the string coupling constant.
The D8-branes and $\overline{\mathrm{D8}}$-branes end at the antipodal points of the $\tau$-circle at the boundary ($u \to \infty$).
In the bulk, they merge smoothly at $u = \ukk$, forming a $\cup$-shaped configuration of the $N_f$ D8-branes.
This geometric connection of the D8- and the $\overline{\mathrm{D8}}$-branes realizes the spontaneous breaking of the chiral symmetry $\mathrm{U}(N_f)_L \times \mathrm{U}(N_f)_R \to \mathrm{U}(N_f)_V$ in the IR.
It is convenient to introduce the coordinates $(y, z)$ defined by
\begin{align}
 y = \sqrt{\frac{u^3 - \ukk^3}{\ukk}} \cos(\tau \mkk), \quad z = \sqrt{\frac{u^3 - \ukk^3}{\ukk}} \sin(\tau \mkk).
\end{align}
In these coordinates, the D8-branes are located at $y=0$ and extend along the $z$-direction, which runs from $-\infty$ to $+\infty$.

\subsection{Effective action of D8-branes}
In the large-$N_c$ limit ($N_f \ll N_c$), the D8-branes are treated as probes on the background geometry \eqref{eq:D4_geometry}.
The dynamics of the flavor gauge field $A_M$ on the D8-branes is governed by the Dirac Born Infeld (DBI) action and the Chern Simons (CS) term. 
In addition, to discuss the CSL phase, it is crucial to include the quark mass term and hence the meson mass term.
This is introduced through the tachyon condensation \cite{Bergman:2007pm,Seki:2010ma,Dhar:2008um,Iatrakis:2010jb} or by adding extra D6-branes \cite{Hashimoto:2008sr, Aharony:2008an}.
Here we employ the latter formalism.
Then the total effective action of the D8-branes is given by
\begin{align}
 S_{\mathrm{D}8} = S_{\mathrm{DBI}} + S_{\mathrm{CS}} + S_{\mathrm{mass}}.
\end{align}
The DBI action, expanded to leading order in $\alpha'$, is given by
\begin{align}
 S_{\mathrm{DBI}} = - \kappa \int d^4x dz \, \tr \left[ \frac{1}{2} K(z)^{-1/3} F_{\mu\nu}^2 + K(z) F_{\mu z}^2 \right], 
 \label{eq:S_DBI}
\end{align}
where the overall constant $\kappa$ is 
given by $ \kappa = \lambda N_c/(216\pi^3)$
and $K=1 + z^2/\ukk^2$.
Here, we have performed the dimensional reduction along the $S^4$ and $A_{6,7,8,9}$ have been set to zero.
The five-dimensional field strength is defined as $F_{MN} = \partial_M A_N - \partial_N A_M + i[A_M, A_N] \ (M,N=\mu,z)$. 
We use the normalization $\tr[\tau^a \tau^b] = 2 \delta^{ab}$ for the gauge generators $T^a$.
Throughout this paper, we adopt the $A_z = 0$ gauge where the gauge field $A_\mu(x, z)$ is expanded in terms of complete sets of mode functions $\psi_n (z)$ as
\begin{align}
 A_\mu(x, z) = A_{L \mu}^{\xi_+} (x) \psi_+ (z) + A_{R \mu}^{\xi_-} (x) \psi_- (z) + \sum_{n=1}^{\infty} B^{(n)}_{\mu} (x) \psi_n (z),
\end{align}
where the zero-mode is given by
\begin{align}
\psi_{\pm} (z) = \frac{1}{2} \Big( 1 \pm \psi_0 (z) \Big),
\qquad
\psi_0 (z) = \frac{2}{\pi} \arctan \left( \frac{z}{\ukk} \right).
\end{align}
We have introduced the following quantities,
\begin{align}
&
A_{L \mu}^{\xi_+} (x) = \xi_+ (x) A_{L \mu} (x) \xi^{-1}_+ (x) - i \xi_+ \del_{\mu} \xi^{-1}_{+} (x),
\notag \\
&
A_{R \mu}^{\xi_-} (x) = \xi_- (x) A_{R \mu} (x) \xi^{-1}_- (x) - i \xi_- \del_{\mu} \xi^{-1}_- (x).
\end{align}
The pion degrees of freedom $\Pi(x)$ are incorporated in 
$\xi_+^{-1} (x) \xi_- (x) = U (x) = e^{2 i \Pi (x)}$ 
while $B_\mu^{(n)}$ are the vector meson towers.
In the following, we neglect the massive vector modes $B^{(n)}_{\mu} = 0$.
Extracting only the zero mode and the second derivative terms in \eqref{eq:S_DBI},
we find the action of the ChPT:
\begin{align}
S_{\mathrm{DBI}} = \int \! d^4 x \,
\frac{f^2}{4} \tr (U^{-1} \del_{\mu} U )^2,
\end{align}
where we have performed the $z$ integration and $f^2 = \frac{1}{54 \pi^4} (g^2_{\mathrm{DBI}} N_c) M^2_{\mathrm{KK}} N_c$ is usually identified with the pion decay constant $f$.

In holographic QCD, the background gauge fields are determined by boundary values as $z \to \infty$.
The boundary condition of the gauge field is given by
\begin{align}
\lim_{z \to \infty} A_{\mu} (x,z) = A^{\xi_+}_{L \mu} (x),
\qquad
\lim_{z \to -\infty} A_{\mu} (x,z) = A^{\xi_-}_{R \mu} (x).
\end{align}
The electromagnetic gauge group $\mathrm{U}(1)_{\mathrm{em}}$ is realized as a part of the gauged chiral symmetry $\mathrm{U}(N_f)_L \times \mathrm{U}(N_f)_R$. 
That is, $\mathrm{SU}(N_f)_L \times \mathrm{SU}(N_f)_R \to \mathrm{SU}(N_f)_V \supset \mathrm{U}(1)_{\mathrm{em}}$.
Here, $SU(N_f)_V$ is the diagonal subgroup of $\mathrm{SU}(N_f)_L \times \mathrm{SU}(N_f)_R$.
On the other hand, the baryon number gauge group $\mathrm{U}(1)_B$ is introduced as a part of $\mathrm{U}(N_f)_V$, namely $\mathrm{U}(N_f)_V = \mathrm{SU}(N_f) \times \mathrm{U}(1)_B$.

The Chern-Simons term plays a central role in our study of the CSL, as it describes the coupling between the baryon density and the external fields such as magnetic fields and rotation. 
It is given by
\begin{align}
S_{\mathrm{CS}} = \frac{N_c}{48 \pi^2} \int_{M^4} \! Z 
+ 
\frac{N_c}{240 \pi^2} \int_{M^4 \times \mathbb{R}} \tr (g d g^{-1})^5
\end{align}
where the second term vanishes for $N_f = 1,2$ which we always assume in this paper.
For the first term, we use the decomposition
\begin{align}
A = A_Q Q + \tilde{A}_B \mathbf{1}
\end{align}
where $Q$ is the electromagnetic charge matrix and $\tilde{A}_B = A_B/N_c$ is the baryon charge gauge field.
Then we obtain
\begin{align}
\frac{N_c}{48 \pi^2} \int Z =& \ 
\frac{N_c}{48 \pi^2} 
\varepsilon^{\mu \nu \rho \sigma} 
\int \! d^4 x \,
\Bigg\{
A_{Q \mu} \tr 
\Big[
Q (L_{\nu} L_{\rho} L_{\sigma} + R_{\nu} R_{\rho} R_{\sigma} )
\Big]
\notag \\
& \qquad \qquad \qquad \qquad \qquad 
- i F_{Q \mu \nu} A_{Q \rho} \tr
\Big[
Q^2 (L_{\sigma} + R_{\sigma})
+ \frac{1}{2} Q U Q \del_{\sigma} U^{\dagger}
- \frac{1}{2} Q U^{\dagger} Q \del_{\sigma} U
\Big]
\Bigg\}
\notag \\
& \ 
- \int \! d^4 x \,
A_{B \mu} 
\,
\left(
-
\frac{1}{24 \pi^2}
\right)
\varepsilon^{\mu \nu \rho \sigma}
\Bigg\{
\tr 
\Big[
L_{\nu} L_{\rho} L_{\sigma}
\Big]
- 
3 i \,
\del_{\nu}
\Big(
A_{Q \rho} \tr 
\Big[
Q (R_{\sigma} + L_{\sigma})
\Big]
\Big)
\Bigg\}
\notag \\
& \ 
- \frac{N_c}{48 \pi^2} \int \! d^4 x \,
\varepsilon^{\mu \nu \rho \sigma} \,
3 i \tilde{A}_{B \mu} \del_{\nu} \tilde{A}_{B \rho} 
\tr \Big[ R_{\sigma} + L_{\sigma} \Big].
\label{eq:WZW_term}
\end{align}
Here $L = U d U^{\dagger}$ and $R = d U^{\dagger} U$.
As we will see in the subsequent sections, the rotation effects are incorporated via the background gauge field in this CS term.
We note that the last term has been neglected in the literature \cite{Son:2007ny}.

Finally, we introduce the extra D6-branes which induce 
the mass term for the meson fields via the worldsheet instanton effects \cite{Hashimoto:2008sr, Aharony:2008an}.
The induced mass term in the four-dimensional effective theory is given by 
\begin{align}
\label{eq:action_pionmass}
 S_{\mathrm{mass}} 
 = 
 \int d^4x \, \frac{m^2 f^2}{4} \tr \Big[ U + U^\dagger - 2 \Big],
\end{align}
where $m$ is the mass parameter determined by the quark mass. 
With this setup, we proceed to analyze the ground state of the system under rotation and magnetic fields.

\section{Rotation as background U(1) gauge field} \label{sec:rotation}
In this section, we first briefly introduce effects of rotation in the gravity dual of QCD.
Details are found in \cite{Amano:2026jsv}.
In analogy with the chemical potential $\mu$, which is identified with the temporal component of a background U(1) gauge field, the angular velocity $\Omega$ can be incorporated through its spatial components of the gauge field $A_i$ \cite{Son:2004tq, Abramchuk:2018jhd}.
This is easily confirmed as follows.
In the four-dimensional rest frame, the chemical potential is incorporated as the temporal component of the background U(1) gauge field $A_\mu=(\mu,0,0,0)$.
The corresponding gauge field configuration in a uniformly rotating frame is obtained by applying a local Lorentz boost with velocity $\vec{v}=\vec{\Omega}\times\vec{x}$, 
where $\vec{\Omega}$ is the constant angular velocity.
A similar approach that introduces rotation via Lorentz boosts in a holographic setup is found in \cite{Zhao:2022uxc, Chen:2020ath}.
Then we have the gauge field $A_\mu =\mu \gamma (v) (1, \vec{v})$ in the boosted frame.
Here $\gamma (v) = 1/\sqrt{1 - v^2}$ is the Lorentz factor.
Assuming that $|v|$ is small $|v| = |x| \Omega \ll c = 1$, 
we have $\gamma \simeq 1 + \mathcal{O}(|v|^2)$ and the 
magnetic field associated with the gauge field is given by
\begin{align}
\vec{B} = \mu \vec{\nabla} \times (\vec{\Omega} \times \vec{x}) = 2 \mu \vec{\Omega}.
\label{eq:rotation_gauge}
\end{align}
Then, the effects of rotation are encoded in the magnetic field of the background U(1) gauge field.
This is consistent with the formalism discussed in \cite{Son:2004tq} 
where rotation in the finite density is described by the background magnetic gauge field $\vec{B} = 2 \mu \vec{\Omega}$.
Note that this identification is true at $\mathcal{O}(|v|)$.

This phenomenon is called the Barnett effect in nuclear matter \cite{Becattini:2020nfu, Becattini:2021lfq, Huang:2017pqe}.
With this identification, we next study the CSL in the holographic setup.

\subsection{CSL in rotation : The \texorpdfstring{$N_f = 1$}{one flavor} case}
We first consider the simplest case of $N_f = 1$ to understand effects of rotation.
Since the rotation is identified as the background U(1) gauge field, it is introduced in the WZW term.
For $N_f = 1$, there is only the $\eta$ meson%
\footnote{
Throughout this paper, we denote the overall U(1) part of $U$ by $\eta$. 
Note that for the $N_f=2$ case, the same U(1) mode was denoted by $\eta'$ in our recent work \cite{Amano:2026jsv}.
}
in the U(1) sector of $U = e^{i \frac{2}{\sqrt{2}} \frac{1}{f} \eta}$.
The baryon number gauge group $\mathrm{U (1)}_{\mathrm{B}}$ is identified with $\mathrm{U(1)}_V$
and we have $L = R = U d U^{\dagger} = - i \frac{1}{f} \sqrt{2} d\eta$
where $f$ is the decay constant associated with $\eta$.
We introduce the rotation as the background magnetic field of the $\mathrm{U(1)}_\mathrm{B}$ baryon gauge field
$\tilde{A}_{B \mu} = \frac{\mu_B}{N_c} (1, \vec{\Omega} \times \vec{x} )$ and hence $\tilde{B}_i = \frac{1}{2} \varepsilon^{ijk} \tilde{F}_{B jk} = \frac{2}{N_c} \mu_B \Omega_i$.
We stress that this is not the magnetic field of the electromagnetic $\mathrm{U(1)}_{\mathrm{em}}$ 
but corresponds to, {\it e.g.} a rotation around the $x^3$ axis (the $x^1$-$x^2$ plane).
The electromagnetic field is introduced later as the Cartan part of
$\mathrm{SU}(N_f)_V$ which will be discussed below.
Setting $Q = 0$ we have the WZW term \eqref{eq:WZW_term}:
\begin{align}
S_{\mathrm{WZW}} =& \ 
- \frac{3N_c}{48 \pi^2} \int \! d^4 x \, 
\varepsilon^{\mu \nu \rho \sigma} \tilde{A}_{B \mu} \tilde{F}_{B \nu \rho} \frac{\sqrt{2}}{f} \del_{\sigma} \eta
\notag \\
=& \ - 
\frac{1}{4 \pi^2}
\frac{1}{N_c}
\int \! d^4 x \, 
\frac{\sqrt{2}}{f} \mu_B^2 \vec{\Omega} \cdot \vec{\nabla} \eta,
\end{align}
where we have assumed $\del_0 \eta = 0$.
Combining with the DBI and the mass parts, the effective action of the D8-branes is given by
\begin{align}
S_{\mathrm{D8}} =& \ \int \! d^4 x \, 
\Bigg[
\frac{f^2}{4} (U^{-1} \del_{\mu} U)^2 
+ 
\frac{m^2 f^2}{4}
(U + U^{-1} - 2)
\Bigg]
+ S_{\mathrm{WZW}}
\notag \\
=& \ 
\int \! d^4 x \, 
\Bigg[
- \frac{1}{2} \del_{\mu} \eta \del^{\mu} \eta 
+ 
\frac{1}{2} m^2 f^2
\Big\{
\cos ( \sqrt{2} f^{-1} \eta) - 1
\Big\}
- \frac{1}{4 \pi^2} \frac{1}{N_c} \frac{\sqrt{2}}{f} \mu_B^2 \vec{\Omega} \cdot
 \vec{\nabla} \eta
\Bigg].
\end{align}
This is the chiral sine-Gordon model for the $N_f = 1$ theory.
Note that the last term does not contribute to the equation of motion.
The equation of motion and the 
solution along the $x^3$-direction are given by
\begin{align}
&
\Box \eta - \frac{m^2 f}{\sqrt{2}} \sin (\sqrt{2} f^{-1} \eta) =
 0, \notag \\
&
\eta (x^3) = \frac{f}{\sqrt{2}} 
\Big\{
\pi 
- 2 \,
\mathrm{am}
(m k^{-1} x^3,k)
\Big\},
\label{eq:N1_amp}
\end{align}
where $\mathrm{am} (x,k)$ is the Jacobi amplitude function and $0 \le k \le 1$ is the elliptic modulus.
For $k=1$, the solution becomes the sine-Gordon kink 
$
\eta (x^3) = 
2 \sqrt{2} f \arctan 
\big[ 
e^{\pm m (x^3 - x^3_0)} 
\big]
$.
The energy density of the solution \eqref{eq:N1_amp} is given by
\begin{align}
\mathcal{E} =& \ \frac{1}{2} (\del_3 \eta)^2 
+
\frac{1}{2} m^2 f^2
\Big\{
1 -  \cos (\sqrt{2}f^{-1} \eta)
\Big\}
- \frac{1}{4 \pi^2} \frac{1}{N_c} \frac{\sqrt{2}}{f} \mu_B^2 \Omega_3 \del_3 \eta
\notag \\
=& \ 
m^2 f^2 
\Bigg\{
k^{-2} \mathrm{dn}^2 (m k^{-1} x^3,k)
+
\mathrm{cn}^2 (m k^{-1} x^3,k)
\Bigg\}
- \frac{1}{4 \pi^2} \frac{1}{N_c} \frac{\sqrt{2}}{f} \mu^2_B \Omega_3
\del_3 \eta,
\end{align}
where we have introduced $\vec{\Omega} = (0,0,-\Omega_3)$.
Then the energy per unit area in the 1-period is evaluated as 
\begin{align}
E =& \ \int_0^{2k \mathcal K(k)/m} \!\!\!  \mathcal{E} dx^3 
\notag \\
=& \ 
2 m f^2 
\left[
\frac{2}{k} E(k) 
+
\left(
k - \frac{1}{k}
\right)
\mathcal{K}(k)
\right]
- \frac{1}{4 \pi} \frac{\sqrt{2}}{N_c} \mu_B^2 \Omega_3,
\end{align}
where $E(k)$ and $\mathcal{K}(k)$ are the complete elliptic integrals of the second and the first kind, respectively.
As in the case of the background magnetic field \cite{Brauner:2016pko}, the ground state is given by the CSL for large-$\Omega_3$.
The energy in the fixed volume $V$ is extremized by $k$ satisfying the following condition:
\begin{align}
\frac{E(k)}{k} = \frac{1}{16 \pi m f^2} \frac{\sqrt{2}}{N_c} \mu_B^2 \Omega_3.
\end{align}
The CSL exists for $\mu_B^2 \Omega_3 \ge \frac{N_c}{\sqrt{2}} 16 \pi m f^2$.
Indeed, it is known that the CSL of the $\eta$ meson becomes the ground state in a rapid rotation 
~\cite{Huang:2017pqe,Nishimura:2020odq,Chen:2021aiq,Eto:2021gyy,Eto:2023tuu,Eto:2023rzd}.

We next examine a brane interpretation of the CSL under rotation.
To this end, we consider the $k=1$ solution for simplicity and evaluate the RR couplings induced by $\eta$ in the D8-brane worldvolume.
The kink induces non-zero gauge field in the D8-branes:
\begin{align}
F_{3z} = - \del_z A_3 = 
\frac{1}{\sqrt{2}} f^{-1}  \del_3 \eta (x^3) \frac{2}{\pi}
 \frac{\ukk^{-1}}{1 + \left( \frac{z}{\ukk}\right)^2}.
 \label{eq:N1_gauge_field_1-kink}
\end{align}
This configuration induces the following RR couplings\footnote{
Although the RR fields have been rescaled as $C^{(p)}
\to \frac{\kappa_{10}^2 \mu_{6-p}}{\pi} C^{(p)}$ in \cite{Sakai:2004cn},
we employ the original form of \cite{Myers:1999ps} in this section.}:
\begin{align}
&
\mu_8 \lambda_s \int \! 
F \wedge C^{(7)}
\notag \\
=& \ 
\mu_8 \lambda_s 
\int_{-\infty}^{\infty} d x^3 
\, 
\frac{1}{\sqrt{2}}
f^{-1}
\del_3 \eta
\int_{-\infty}^{\infty}
dz \,
 \frac{2}{\pi}
 \frac{\ukk^{-1}}{1 + \left(
 \frac{z}{\ukk}\right)^2}
\varepsilon^{3z a_1 \cdots a_7} \frac{1}{7!} C^{(7)}_{a_1 \ldots a_7}
\notag \\
=& \ \pm \mu_6 \, \varepsilon^{3z a_1 \cdots a_7} \frac{1}{7!} C^{(7)}_{a_1 \ldots
 a_7}
\quad (a_1, \ldots, a_7 = 0,1,2,6,7,8,9),
\label{eq:kinkN1RR}
\end{align}
where we have presented only the relevant integrations.
Here $\lambda_s = 2 \pi \alpha'$ and 
$\mu_p = \frac{1}{(2 \pi)^p} g_s^{-1} \alpha^{\prime - (p+1)/2}$ is the D$p$-brane charge 
and $\pm$ corresponds to the kink ($+$) or the anti-kink ($-$).
Therefore, the sine-Gordon kink ($\eta$ soliton) in the absence of rotation is
identified with the D6(0126789)-brane.
Here the notation stands for the D6-brane extending along the $(x^0,x^1,x^2,x^6, \ldots,x^9)$-directions.
Naively, a configuration of D8- and D6-branes with $\sharp_{\mathrm{ND}}=2$ Neumann-Dirichlet  directions is unstable due to the presence of open string tachyons.
However, in our holographic framework, the D6-brane charge is not introduced as an independent probe but is instead realized as a topological kink ($F_{3z} \not= 0$) of the massive meson field on the D8-brane worldvolume. 
Indeed, the QCD domain wall in the Sakai-Sugimoto model with massive quarks has been discussed as a stable D8/D6-brane system
\cite{Argurio:2018uup}.
It has been shown that the mass of the D8-D6 open strings gets
a positive shift as quark masses are increased.

Next, we introduce rotation around the $x^3$ axis (or the $x^1$-$x^2$ plane).
This corresponds to turning on the constant baryon magnetic field
$F_{12} = 2 N_c^{-1} \mu_B \Omega_3$
in the D8-brane.
This induces a non-zero RR coupling of the form
\begin{align}
&
\mu_8 \lambda_s \int 
 F \wedge C^{(7)}
\notag \\
=& \ \mu_6 
\int \! dx^1 dx^2 \, \frac{2 N_c^{-1} \mu_B \Omega_3}{2\pi} 
\, 
\varepsilon^{12 a_1 \cdots a_7}
\frac{1}{7!} C^{(7)}_{a_1 \cdots a_7}
\notag \\
=& \  \mu_6 \mathbf{\Omega}_3 \,  \varepsilon^{12 a_1 \cdots a_7}
\frac{1}{7!} C^{(7)}_{a_1 \cdots a_7} \quad (a_1, \ldots, a_7 = 0,3,z,6,7,8,9),
\label{eq:B3N1RR}
\end{align}
where $\mathbf{\Omega}_3 = \int \! dx^1 dx^2 \, \frac{\mu_B \Omega_3}{N_c \pi}$ is the ``flux'' in the
$(x^1,x^2)$-plane.
When this value is an integer $n$, the RR coupling carries $n$ units of D6$(03z6789)$-brane charge.
In the rotating configuration considered here, however, the flux need not be quantized. 
An appropriate interpretation is therefore a continuous D6 charge density spread out and uniformly distributed in the $(x^1,x^2)$-plane, namely, the object is the dissolved D6-branes \cite{Zwiebach:2004tj}.

\begin{figure}[t]
\begin{center}
\includegraphics[scale=1]{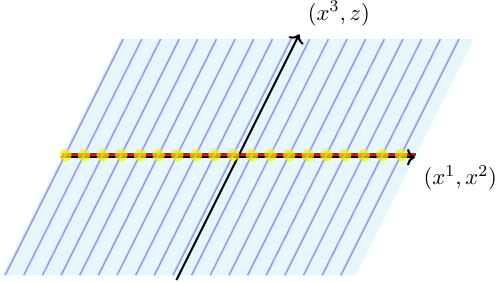}
\end{center}
\caption{
A schematic picture of the brane configuration for the kink in the
 rotation ($N_f = 1$).
The blue shaded region represents the D8-brane worldvolume
that fills the whole $(x^1,x^2,x^3,z)$-plane.
The blue vertical lines are the $\mu_B \Omega$-induced D6-branes extending along the $(x^3,z)$-directions and uniformly distributed along the $(x^1,x^2)$-directions.
The red horizontal line is the kink D6-brane extending along the
 $(x^1,x^2)$-directions and localized in the $(x^3,z)$-plane.
The yellow dots are the D4-branes localized on the $(x^3,z)$-directions and
 uniformly distributed along the $(x^1,x^2)$-directions.
The D4-branes appear at the intersections of the two types of D6-branes.
The D6-branes for the mass deformation and the $(x^0,x^6,x^7,x^8,x^9)$-directions are omitted in the figure.
}
\label{fig:D8D6D4}
\end{figure}

Now we consider the sine-Gordon kink in the rotation.
In this case, we have the non-zero gauge field components in the D8-branes,
\begin{align}
F_{12} = 2 N_c^{-1} \mu_B \Omega_3,
\qquad
F_{3z} = \frac{1}{\sqrt{2}} f^{-1} \del_3 \eta (x^3) \frac{2}{\pi}
 \frac{\ukk^{-1}}{1 + \left( \frac{z}{\ukk}\right)^2}.
\end{align}
In addition to the couplings \eqref{eq:kinkN1RR} and \eqref{eq:B3N1RR}, 
the gauge field induces the following RR coupling:
\begin{align}
&
\mu_8 \frac{\lambda_s^2}{2} 
\int 
F \wedge F \wedge C^{(5)}
\notag \\
=& \ 
\mu_8 \lambda_s^2 \int \! dx^1 dx^2 \,
\varepsilon^{123z a_1 \cdots a_5}
F_{12} F_{3z} 
\frac{1}{5!} C^{(5)}_{a_1 \cdots a_5}
\notag \\
\label{eq:action_F2}
=& \ 
\pm \mu_4 \mathbf{\Omega}_3 \, \varepsilon^{123z a_1 \cdots a_5} 
\frac{1}{5!} C^{(5)}_{a_1 \cdots a_5}, \quad (a_1, \ldots, a_5 = 0,6,7,8,9).
\end{align}
In this case, we find that the configuration induces the D4(06789)-charge.
The D4-brane charge emerges at the intersection of the kink D6(0126789) localized
in the $(x^3,z)$-plane and the $\mu_B \Omega$-induced D6(03$z$6789) dissolved in the $(x^1,x^2)$-plane.
Therefore, the kink in the rotation looks like a partly dissolved D4$(06789)$-brane 
localized in the $(x^3,z)$-plane and distributed in the $(x^1,x^2)$-plane
(see Table~\ref{tb:dissolved-D4} and Figure~\ref{fig:D8D6D4}).

\begin{table}[t]
\begin{center}
\begin{tabular}{c|c|c|c|c|c|c|c|c|c|c}
 & 0 & 1 & 2 & 3 & $y$ & $z$ & 6 & 7 & 8 & 9 \\
\hline
\hline
$N_f$ D8 & $\circ$ & $\circ$ & $\circ$ & $\circ$ &  & $\circ$ & $\circ$ & $\circ$ & $\circ$ & $\circ$ \\
\hline
\hline
kink $=$ D6
& 
$\circ$ 
& 
$\circ$ 
& 
$\circ$  
& 
& 
& 
& 
$\circ$ 
& 
$\circ$ 
& 
$\circ$ 
& 
$\circ$ 
 \\
\hline
$\mu_B \Omega$ $=$ D6
& 
$\circ$ 
&
\cellcolor{gray!20}
&
\cellcolor{gray!20}
& 
$\circ$
& 
& 
$\circ$  
& 
$\circ$ 
& 
$\circ$ 
& 
$\circ$ 
& 
$\circ$ 
 \\
\hline
\hline
(kink + $\mu_B \Omega$) $=$ D4
& 
$\circ$ 
& 
\cellcolor{gray!20}
& 
\cellcolor{gray!20}
& 
& 
& 
& 
$\circ$ 
& 
$\circ$ 
& 
$\circ$ 
& 
$\circ$ 
 \\
\end{tabular}
\end{center}
\caption{The sine-Gordon kink ($N_f=1$) together with the constant
 angular velocity $\mu_B \Omega_3$ induces non-zero
 RR-charges for D6(0126789), D6$(03z6789)$ and D4(06789).
The symbol $\circ$ stands for the worldvolume directions while the colored cell indicates that the D-branes are dissolved in this direction.
}
\label{tb:dissolved-D4}
\end{table}

This is similar to the CSL in a magnetic field \cite{Amano:2025iwi}. 
However, in that case, the charge of the D6-branes is canceled, leaving only the charge of the D4-branes. 
It is interesting that in the case of rotation, the charge of the D6-branes remains along with that of the D4-branes.

\subsection{CSL in rotation and magnetic field : The \texorpdfstring{$N_f = 2$}{two flavors} case}
\label{sec:chpt_csl_nf2}

We next study the $N_f=2$ case.
In this case, we can introduce the background magnetic field together with rotation.
We focus on the neutral pion $\pi^0$ and the eta meson $\eta$ in this section and ignore charged pions.
This means that we have the meson field\footnote{
Note that we have introduced the common decay constants for the pion and the eta sectors $f_{\pi} = f_{\eta} = f$ since there is only a single decay constant $f$ in our setup.
}
$U = \exp \left[i \left( \tau^3 \frac{\pi^0}{f} + \tau^0 \frac{\eta}{f} \right) \right]$ where $\tau^0 \equiv \mathbf{1}_2$.
We introduce the electromagnetic field as the Cartan part of $\mathrm{SU}(2)_V$.
The electromagnetic charge is defined by $$Q = \frac{\tau^3}{2} + \frac{1}{6} \tau^0\,.$$
Each part of the gauge field $A = A_Q Q + \frac{A_B}{N_c} \tau^0$ is then given by
\begin{align}
&
A_{Q0} = 0,
\qquad
A_{Qi} = A_{Qi} (\vec{x}),
\qquad
F_{Qij} = \varepsilon_{ijk} B_k = \text{const.},
\notag \\
&
A_{B0} = \mu_B,
\qquad
A_{Bi} = A_{Bi} (\vec{x}),
\qquad
F_{Bij} = 2 \mu_B \varepsilon_{ijk} \Omega_k = \text{const.}
\end{align}
Namely, the background gauge field is given by
\begin{align}
&
A_0 = \frac{1}{N_c} \mu_B \tau^0,
\qquad
A_i = A_{Qi} 
\left(
\frac{\tau^3}{2} + \frac{1}{6} \tau^0
\right)
+
\frac{A_{Bi}}{N_c} \tau^0,
\notag \\
&
F_{ij} = 
\varepsilon_{ijk}
\left\{
B_k 
\left(
\frac{\tau^3}{2} + \frac{1}{6} \tau^0
\right)
+
\frac{2 \mu_B \Omega_k}{N_c} \tau^0
\right\}\,.
\end{align}
Then we find that the action of the D8-branes becomes
\begin{align}
S_{D8} =& \
\int \! d^4 x \,
\Bigg[
- \frac{1}{2} \del_{\mu} \pi^0 \del^{\mu} \pi^0
- \frac{1}{2} \del_{\mu} \eta \del^{\mu} \eta
+  m^2 f^2 
\left\{
\cos \left( \frac{\eta}{f} \right) 
\cos \left( \frac{\pi^0}{f} \right)
-1
\right\}
\notag \\
& \qquad \qquad
-
\frac{1}{4 \pi^2} \frac{1}{f}
\mu_B B_k
\left(
\del_k \pi^0 + \frac{1}{3} \del_k \eta
\right)
-
\frac{1}{2 \pi^2} \frac{1}{f} \frac{1}{N_c} \mu_B^2 
\Omega_k \del_k \eta
\Bigg].
\end{align}
Again, the second line does not contribute to the equation of motion.

The equations of motion are given by
\begin{align}
& 
\Box \eta - m^2 f \sin \left( \frac{\eta}{f} \right)
\cos \left( \frac{\pi^0}{f} \right)  
= 0,
\notag \\
&
 \Box \pi^0 - m^2 f 
 \cos \left( \frac{\eta}{f} \right)
 \sin \left( \frac{\pi^0}{f} \right)  = 0.
\end{align}
Defining $\phi_{\pm} = \frac{1}{f} (\eta \pm \pi^0)$, the equations become \cite{Eto:2021gyy}
\begin{align}
\label{eq:chtp_csl_eom}
\Box \phi_{\pm} - m^2 \sin \phi_{\pm} = 0.
\end{align}
Assuming that $\phi_{\pm}$ depends only on $x^3$, 
the solution to this equation is given by
\begin{equation}
\label{eq:phi_pm_amplitude}
\begin{aligned}
\phi_+ (x^3) =& \ \pi \pm 2 \, \mathrm{am} \left(\frac m{k_+} (x^3-x_0^\pm), k_+\right), \\
\phi_- (x^3) =& \ \pi \pm 2 \, \mathrm{am} \left(\frac m{k_-} (x^3-\tilde x_0^\pm), k_-\right), \\
\end{aligned}
\end{equation}
where $\mathrm{am} (x,k)$ is the Jacobi amplitude function and $k_{\pm}$ are the elliptic moduli.
For $k_{\pm} = 1$, we have 1-kink (1-antikink) solutions,
\begin{align}
\phi_{\pm} = 
4 \, \arctan \Big( e^{m (x^3 - x_0^{\pm})} \Big),
\quad
4 \, \arctan \Big( e^{- m (x^3-\tilde x_0^\pm)} \Big).
\label{eq:phi_pm_kink}
\end{align}
The energy density $\mathcal{E}$ for the solution \eqref{eq:phi_pm_amplitude} is given by
\begin{align}
\label{eq:chpt_energy_density}
\mathcal{E} =& \ 
\frac{1}{2} (\del_3 \pi^0)^2
+
\frac{1}{2} (\del_3 \eta)^2
+ 
m^2 f^2 
\left\{
1 - 
\cos \left( \frac{\eta}{f} \right) 
\cos \left( \frac{\pi^0}{f} \right)
\right\}
\notag \\
& \ 
- \frac{1}{4 \pi^2} \frac{1}{f}
\mu_B B_3
\left(
\del_3 \pi^0 + \frac{1}{3} \del_3 \eta
\right)
- \frac{1}{2 \pi^2} \frac{1}{f} \frac{1}{N_c} \mu_B^2 
\Omega_3 \del_3 \eta
\notag \\
=& \ 
\frac{f^2}{4} (\del_3 \phi_+)^2
+ \frac{m^2 f^2}{2} 
\Big(
1 - \cos \phi_+
\Big)
- \frac{\mu_B}{4 \pi^2} 
\left(
\frac{2}{3} B_3 +  \frac{2}{N_c} \mu_B \Omega_3
\right) \del_3 \phi_+
\notag \\
+ & \
\frac{f^2}{4} (\del_3 \phi_-)^2
+ \frac{m^2 f^2}{2} 
\Big(
1 - \cos \phi_- 
\Big)
- \frac{\mu_B}{4 \pi^2}
\left(
- \frac{1}{3} B_3 + \frac{2}{N_c} \mu_B \Omega_3
\right) \del_3 \phi_-,
\end{align}
where we have introduced $\vec{B} = (0,0,-B_3)$ and $\vec{\Omega} = (0,0,-\Omega_3)$.
Note that the $\phi_+$ and $\phi_-$ parts are completely decoupled 
$\mathcal{E} = \mathcal{E}_+ + \mathcal{E}_-$.
Here $\mathcal{E}_{\pm}$ corresponds to terms including $\phi_{\pm}$.

\begin{figure}[t]
  \centering
  \includegraphics[width=0.75\textwidth]{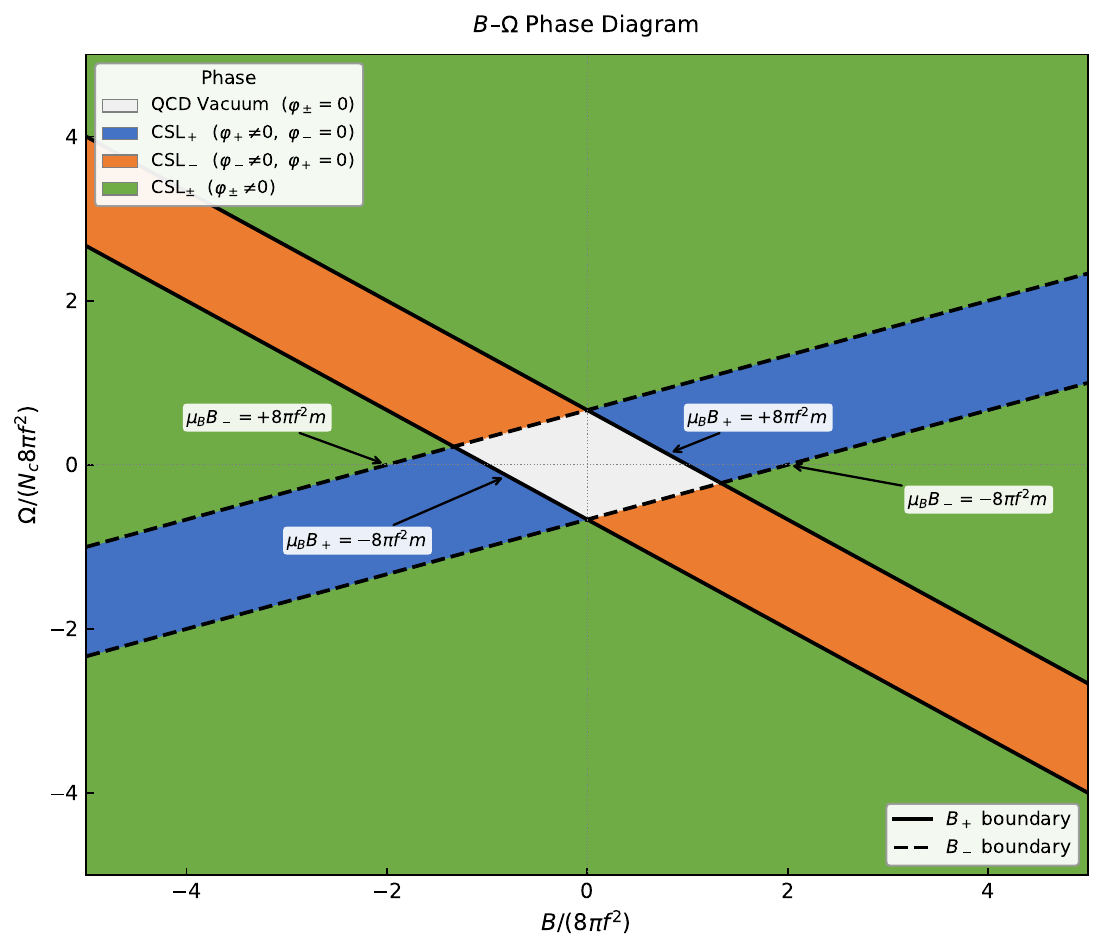}
  \caption{Phase diagram of the $N_f=2$ ChPT CSL in the $B$--$\Omega$ plane
    at fixed baryon chemical potential $\mu_B = 1.5\,(8\pi f^2)^{1/2}$ where $m=1$.
    The four phases are separated by the boundaries
    $\mu_B|B_\pm| = 8\pi f^2 m$, where
    $B_+ = \frac{2}{3}B + \frac{2\mu_B}{N_c}\Omega$ and
    $B_- = -\frac{1}{3}B + \frac{2\mu_B}{N_c}\Omega$.
    Solid (dashed) lines denote $B_+$ ($B_-$) boundaries.}
  \label{fig:phase_B_Omega}
\end{figure}

\begin{figure*}[t]
  \centering
  \includegraphics[width=0.33\textwidth]{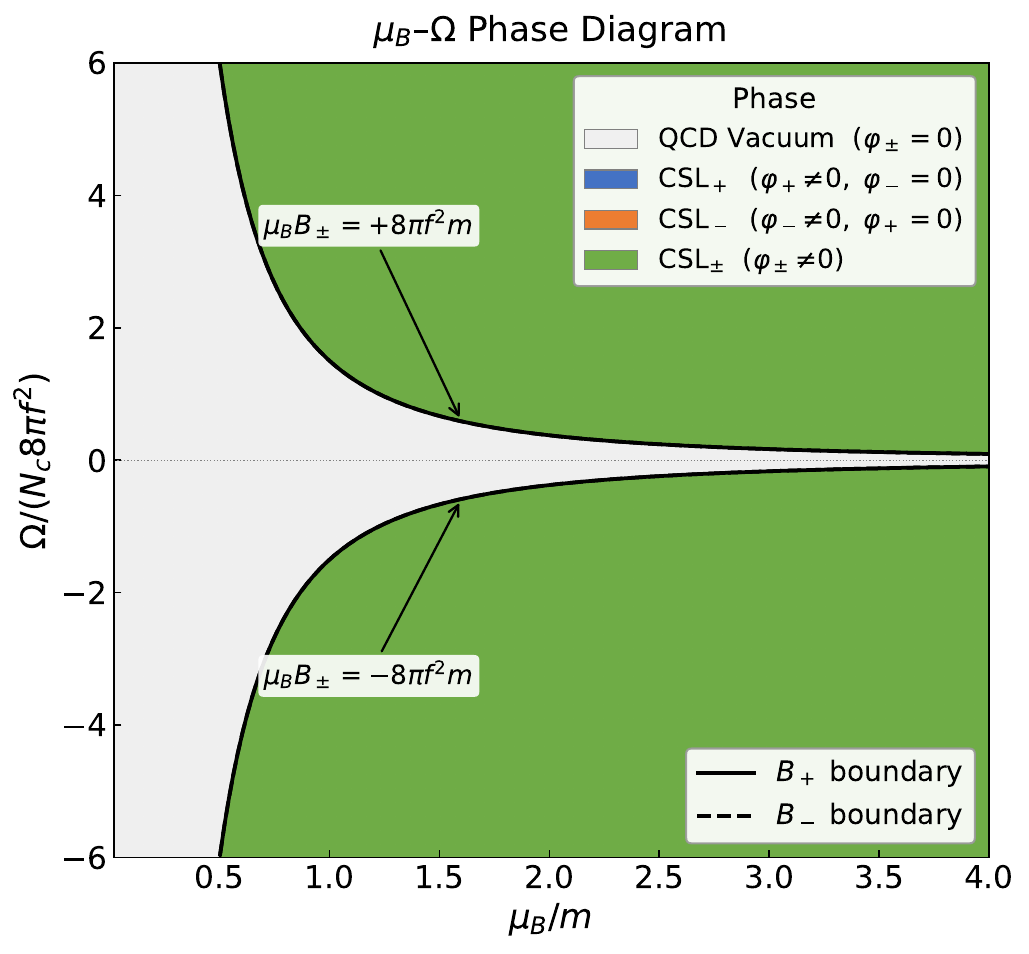}%
  \includegraphics[width=0.33\textwidth]{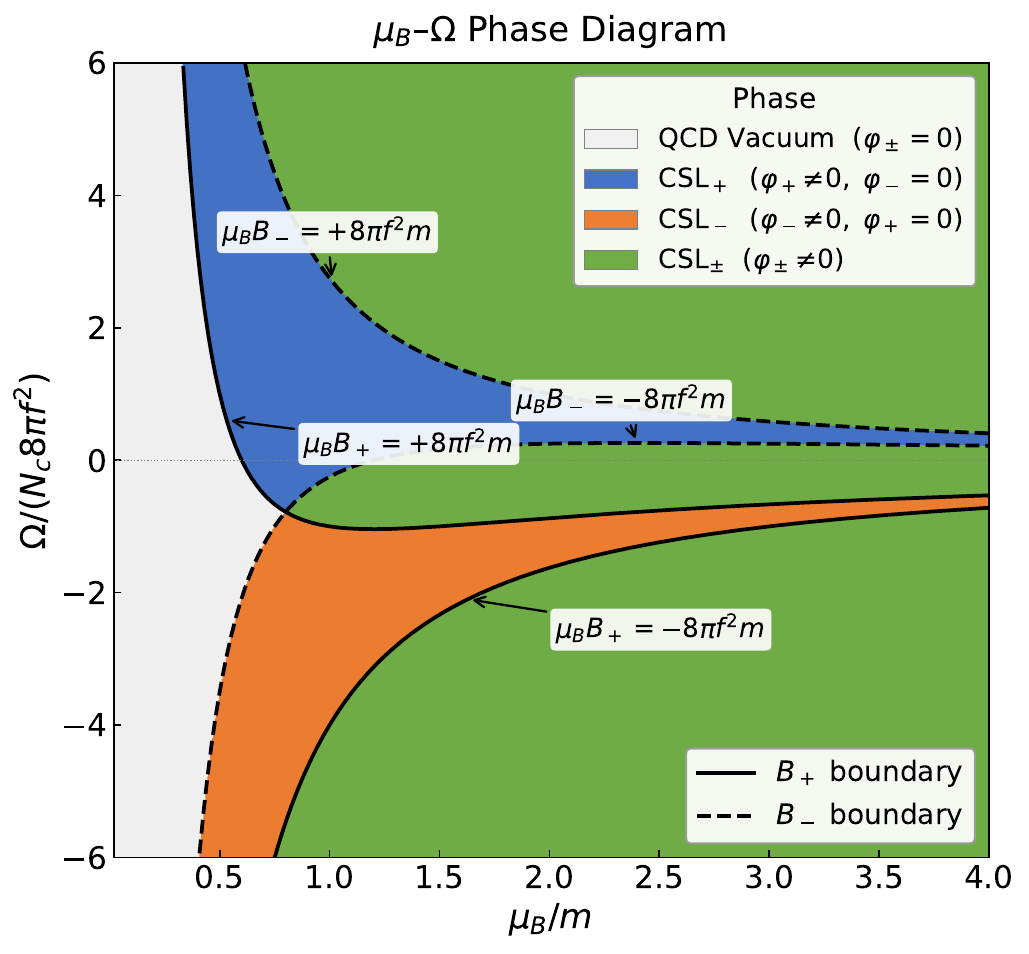}%
  \includegraphics[width=0.33\textwidth]{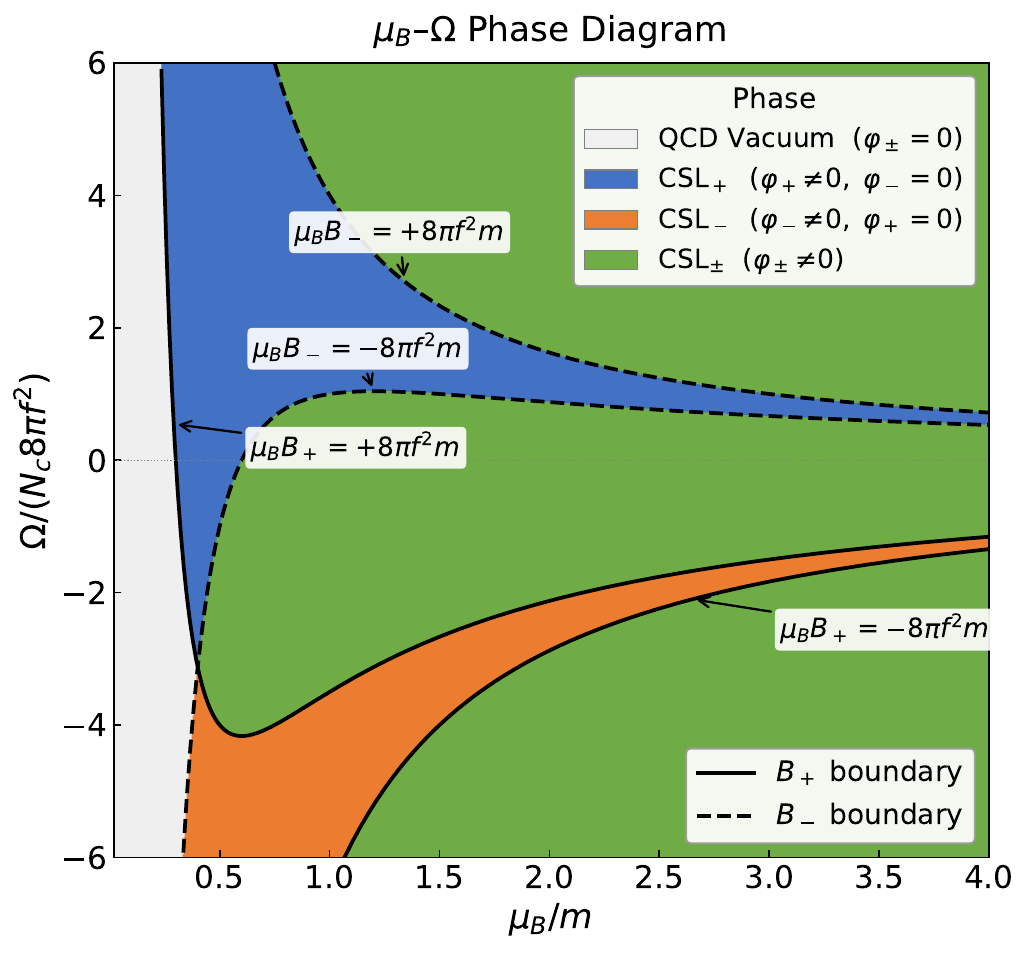}
  \caption{Phase diagrams in the $\mu_B/m$--$\Omega/(N_c 8\pi f^2)$ plane.
    Left: $B/(8\pi f^2)=0$, where $B_+ = B_-$ and the two boundaries are degenerate,
    leaving only the QCD vacuum and $\mathrm{CSL}_{\pm}$ phases.
    Middle: $B/(8\pi f^2)=2.5$ and Right: $B/(8\pi f^2)=5.0$, where the degeneracy is lifted
    and all four phases appear, with the $\mathrm{CSL}_-$ band widening as $B$ increases.
    Solid (dashed) lines denote $B_+$ ($B_-$) boundaries
    $(\mu_B/m)\,|B_\pm|/(8\pi f^2) = 1$.}
  \label{fig:phase_mu_Omega}
\end{figure*}

\begin{figure*}[t]
  \centering
  \includegraphics[width=0.33\textwidth]{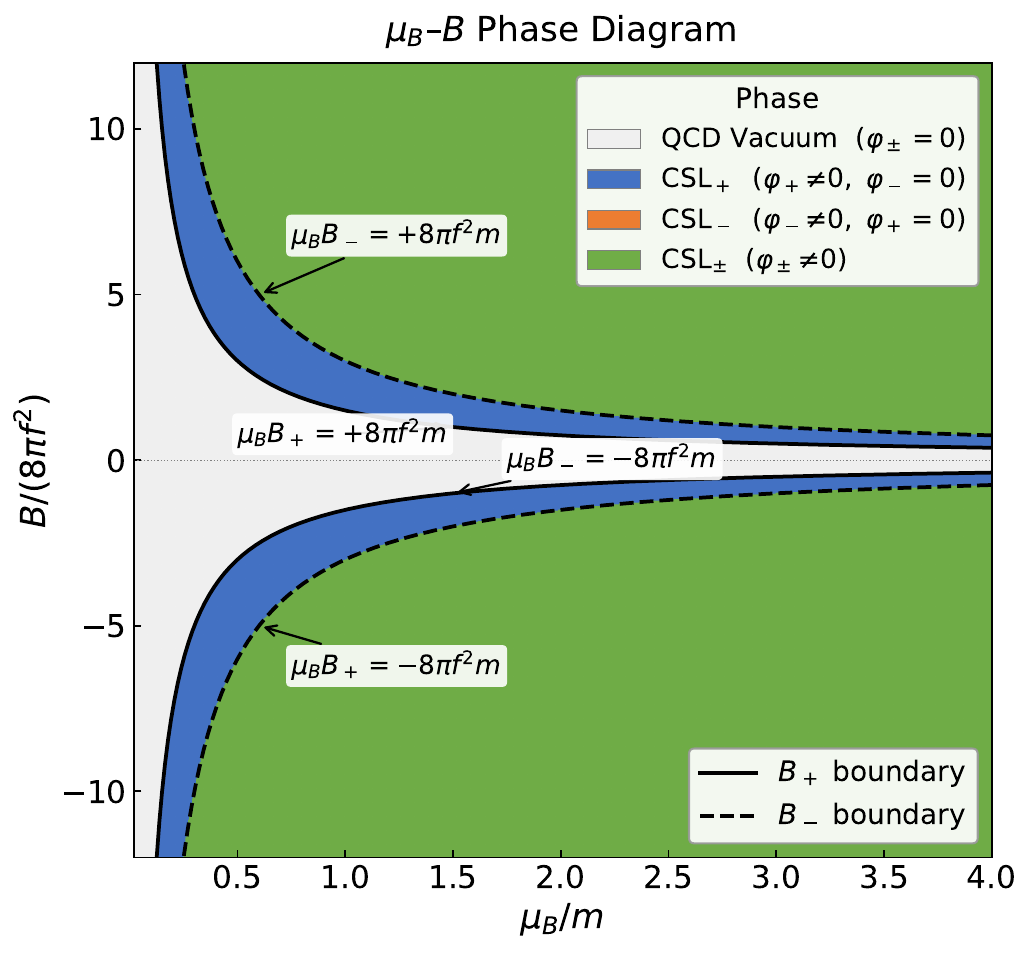}%
  \includegraphics[width=0.33\textwidth]{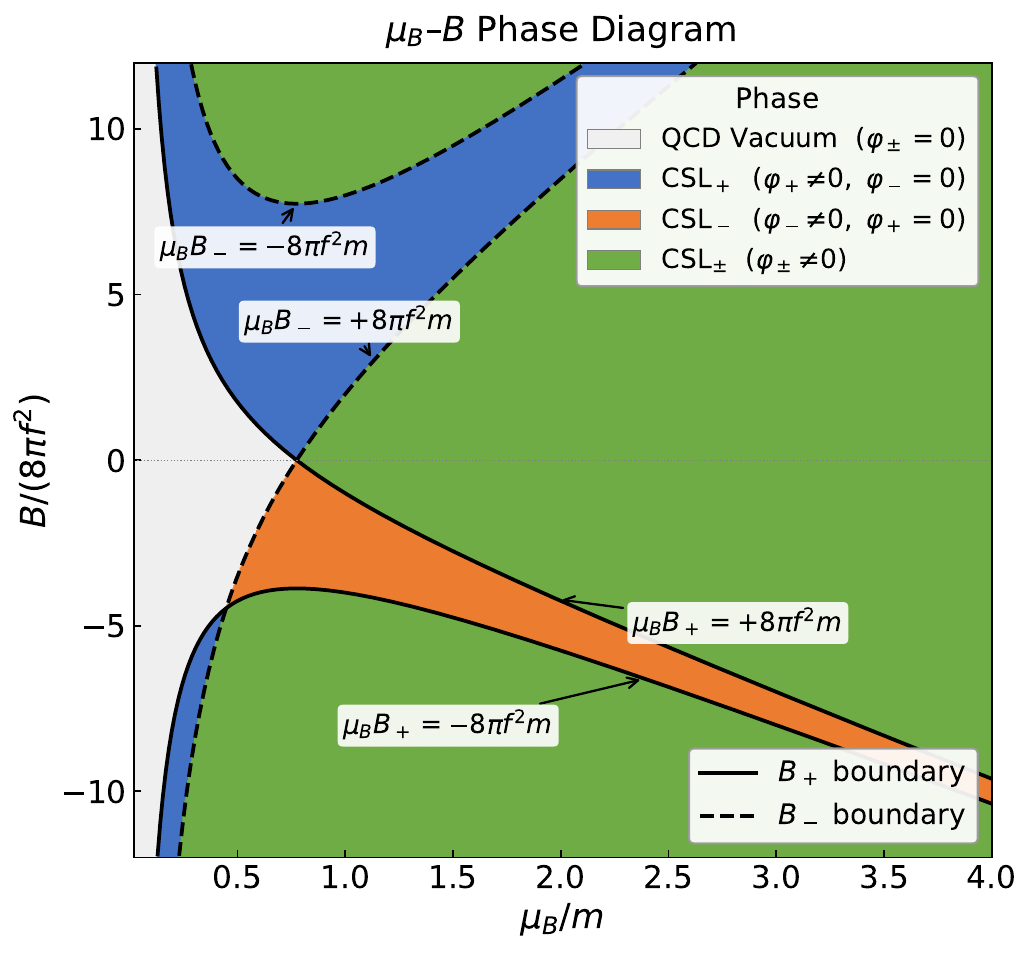}%
  \includegraphics[width=0.33\textwidth]{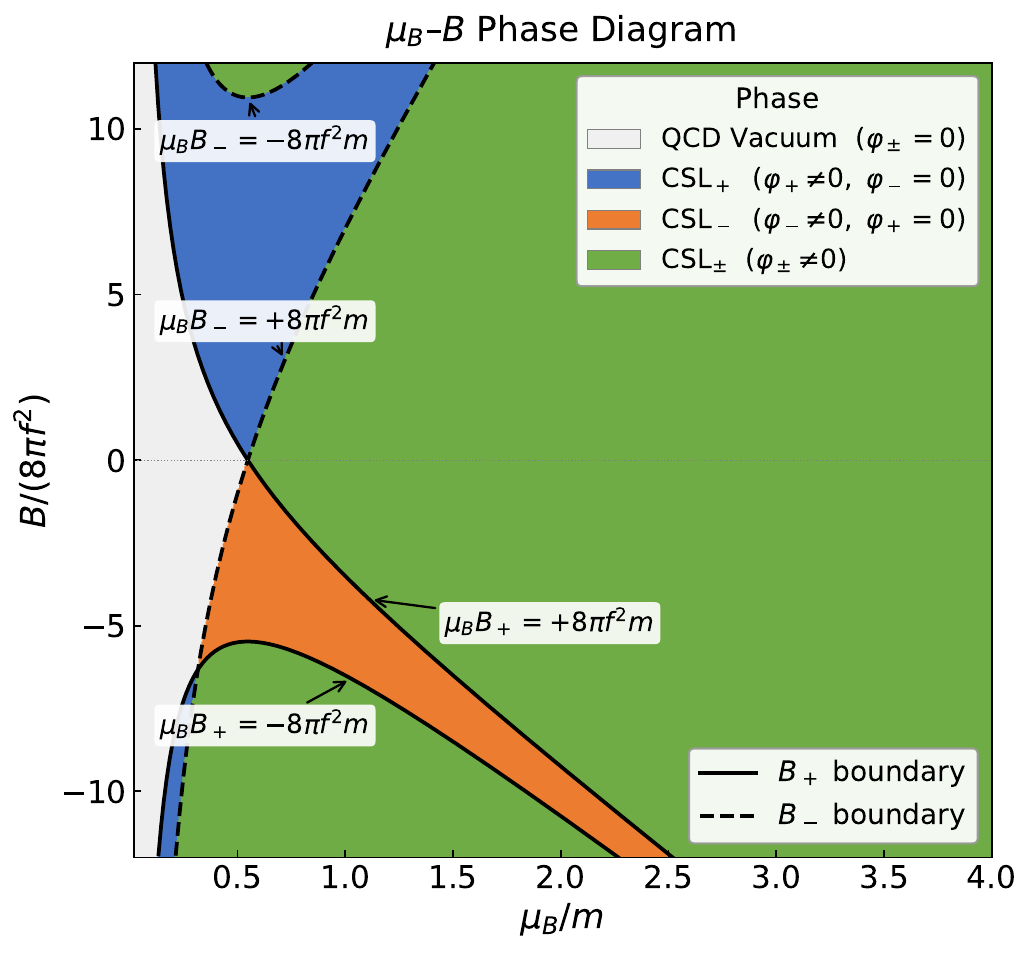}
  \caption{Phase diagrams in the $\mu_B/m$--$B/(8\pi f^2)$ plane.
    Left: $\Omega/(N_c 8\pi f^2)=0$, where no $\mathrm{CSL}_-$ phase appears since
    $|B_+| \geq |B_-|$ throughout.
    Middle: $\Omega/(N_c 8\pi f^2)=2.5$ and Right: $\Omega/(N_c 8\pi f^2)=5.0$, where rotation
    breaks this inequality and a $\mathrm{CSL}_-$ band emerges at negative~$B$, widening as
    $\Omega$ increases.
    Solid (dashed) lines denote $B_+$ ($B_-$) boundaries
    $(\mu_B/m)\,|B_\pm|/(8\pi f^2) = 1$.}
  \label{fig:phase_mu_B}
\end{figure*}

Looking at the decoupled sectors, we can frame it as two energy minimization problems.
Each integrand takes the following form
\begin{equation}
    \label{eq:chpt_off_shell_energy}
\mathcal{E}_{\pm} = \frac{f^2}{4} (\del_3 \phi_\pm)^2 - \frac{m^2 f^2}{2} \cos \phi_\pm -
    \frac{\mu_B}{4\pi^2} B_\pm \del_3 \phi_\pm
    + \frac{1}{2} m^2 f^2
\end{equation}
where we have defined
\begin{equation}
    \begin{aligned}
        B_+ &=&&\frac{2}{3} B_3 + \frac{2}{N_c} \mu_B \Omega_3, \\
        B_- &=&-&\frac{1}{3} B_3 + \frac{2}{N_c} \mu_B \Omega_3. \\
    \end{aligned}
\end{equation}
Note that we have `split' the $m^2 f^2$ term since it does not depend on $\phi_\pm$ and that the summation of both of these energies recovers \eqref{eq:chpt_energy_density}.
One can see this separation as needed because for the $\phi_\pm = \mathrm{const}$ case, we could always choose $\phi_\pm = 0$, so its individual contribution in the mass term vanishes.
Let us now substitute the on-shell solution of \eqref{eq:phi_pm_amplitude}, integrating and averaging each period:
\begin{equation}
    \sigma_\pm := \frac{1}{T_\pm}\int dx^3 \mathcal E(k_\pm) =  \frac{m \left(4 \pi  f^2 m \left(\left(k_\pm^2-1\right) \mathcal K(k_\pm)+2 E(k_\pm)\right)-B_\pm k_\pm \mu_B\right)}{4 \pi  k_\pm^2 \mathcal K(k_\pm)},
\end{equation}
where $T_{\pm}$ is the period of each $\pm$ sector.
Varying with respect to $k_\pm$ and setting to zero we find that
\begin{equation}
    \frac{\pd \sigma}{\pd k_\pm} = 0 \implies \frac{E(k_\pm) \left(8 \pi f^2 m E(k_\pm) - B_\pm k_\pm \, \mu_B\right)}{k_\pm} = 0 \implies
    \mu_B B_\pm = 8 \pi f^2 m \frac{E(k_\pm)}{k_\pm}.
\end{equation}
This implies a condition on $B_\pm$.
Since $E(k_\pm)/k_\pm > 1$ if $0 < k_{\pm} < 1$, we can see that
\begin{equation}
    \mu_B B_\pm \geq 8 \pi f^2 m.
\end{equation}
Accounting for the $\pm$ sign choice in \eqref{eq:phi_pm_amplitude}, the condition becomes
\begin{equation}
    \label{eq:chpt_bpm_csl_condition}
    \mu_B |B_\pm| \geq 8 \pi f^2 m
\end{equation}
where a positive $B_\pm$ induces a CSL while a negative $B_\pm$ induces an anti-CSL.
Substituting back in $B_Q$ and $\Omega_3$ we have 
\begin{equation}
    \begin{aligned}
        \mu_B \left|\frac{2}{3} B_3 + \frac{2}{N_c} \mu_B \Omega_3\right| &\geq 8 \pi f^2 m, \\
        \mu_B \left|-\frac{1}{3} B_3 + \frac{2}{N_c} \mu_B \Omega_3\right| &\geq 8 \pi f^2 m. \\
    \end{aligned}
\end{equation}
When either condition is not met, the preferred state for that sector is the QCD vacuum, $\phi_\pm = 0$.
When the condition is met, the corresponding field forms a CSL.
The four resulting phases, including the trivial QCD vacuum, are summarized in Table~\ref{tab:chpt_csl_phases}.
In the single-CSL phases, a residual $\mathrm{U}(1)$ symmetry locks $\eta$ and $\pi^0$ to be perfectly in or out of phase, i.e.\ $\eta = \pm\,\pi^0$.
Phase diagrams can be seen in Figure \ref{fig:phase_B_Omega}, Figure \ref{fig:phase_mu_Omega}, and Figure \ref{fig:phase_mu_B}.

\begin{table}[t]
\centering
\renewcommand{\arraystretch}{1.6}
\begin{tabular}{l c c c c}
\hline\hline
Phase 
  & $\mu_B|B_+|$
  & $\mu_B|B_-|$
  & $\eta(x^3)$ 
  & $\pi^0(x^3)$ \\
\hline
QCD vacuum 
  & ${}<8\pi f^2 m$ & ${}<8\pi f^2 m$ 
  & $0$ 
  & $0$ \\[4pt]
$\phi_+$-CSL (aligned) 
  & $\geq8\pi f^2 m$ & ${}<8\pi f^2 m$ 
  & $\dfrac{f}{2}\,\phi_+$ 
  & $\dfrac{f}{2}\,\phi_+$ \\[4pt]
$\phi_-$-CSL (anti-aligned) 
  & ${}<8\pi f^2 m$ & $\geq8\pi f^2 m$ 
  & $\dfrac{f}{2}\,\phi_-$ 
  & $-\dfrac{f}{2}\,\phi_-$ \\[4pt]
Dual CSL 
  & $\geq8\pi f^2 m$ & $\geq8\pi f^2 m$ 
  & $\dfrac{f}{2}\!\left(\phi_+ + \phi_-\right)$ 
  & $\dfrac{f}{2}\!\left(\phi_+ - \phi_-\right)$ \\
\hline\hline
\end{tabular}
\caption{
Phase structure of the $\eta$--$\pi^0$ system.  
}
\label{tab:chpt_csl_phases}
\end{table}

\begin{figure}[t]
\begin{center}
\includegraphics[scale=0.5]{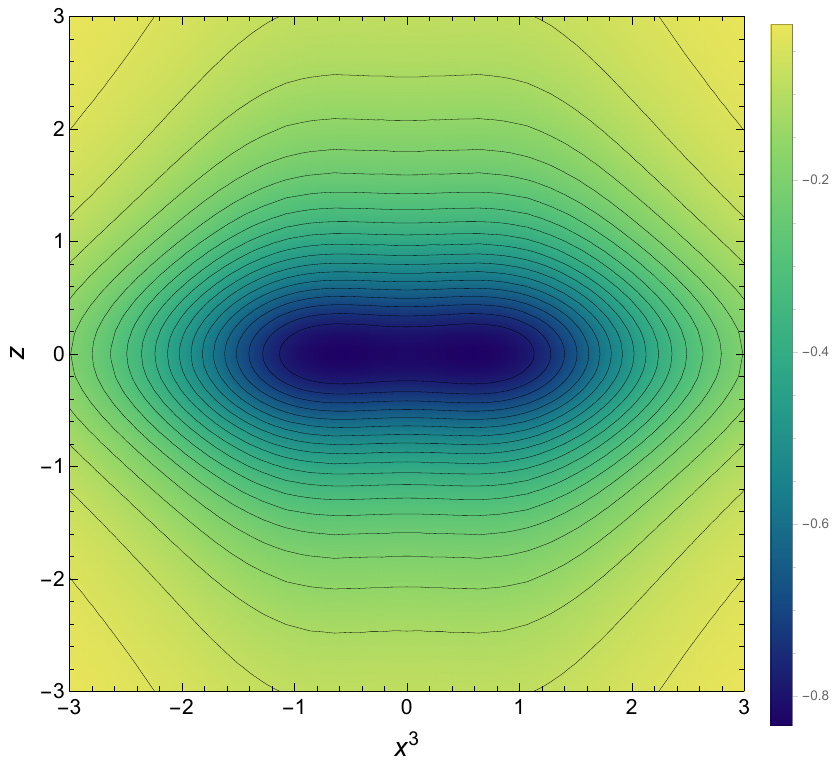}
\includegraphics[scale=0.5]{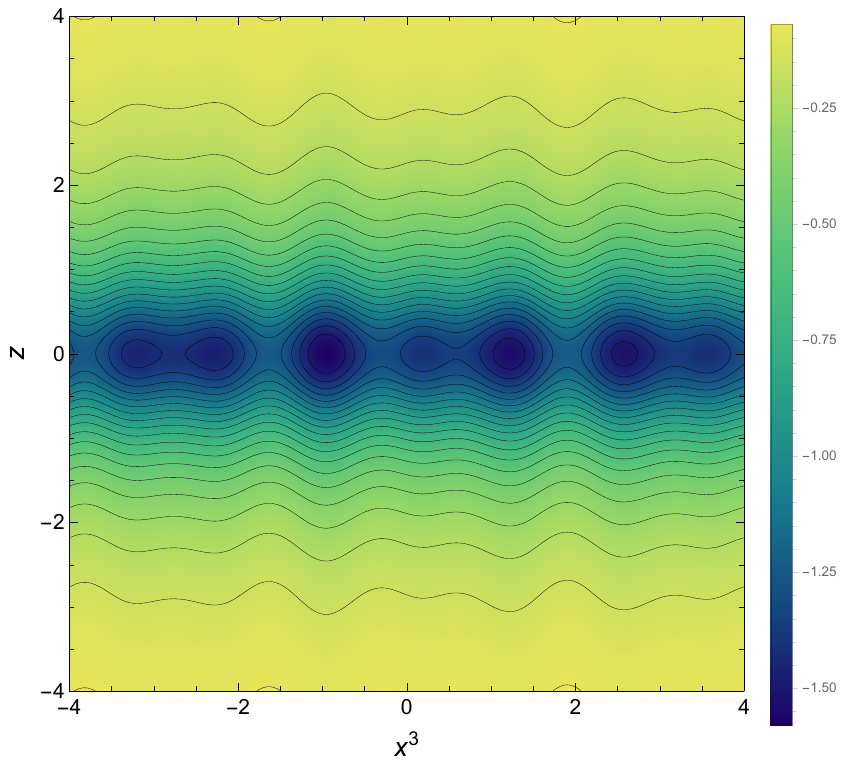}
\end{center}
\caption{
The vortex density $\tr [F_{3z}]$ for $\eta$-$\pi^0$ CSL.
Left: 1-kink for $\phi_+, \phi_-$ where the kink positions are set to $x^+_0 = -1, \tilde x^+_0 = 1$.
Right: CSL for the solution \eqref{eq:phi_pm_amplitude} where $k_+ = 1/2, k_- = 1/3$.
Parameters $\mu_B, m, B_3, \Omega_3, \ukk, N_c$ are set to be 1.
}
\label{fig:eta_CSL}
\end{figure}

\paragraph{Brane interpretation of CSL}
We next examine the RR charges induced by the CSL and the backgrounds.
The non-zero components of the gauge field strength on the D8-brane worldvolume are evaluated as
\begin{align}
F_{12} =& \ 
B_3 \left( \frac{\tau^3}{2} + \frac{1}{6} \tau^0 \right)
+ \frac{2}{N_c} \mu_B \Omega_3 \tau^0,
\notag \\
F_{3z} 
=& \ - f^{-1}
\left(
\frac{\tau^3}{2} \del_3 \pi^0
+
\frac{\tau^0}{2} \del_3 \eta
\right)
\frac{2}{\pi}
\frac{\ukk^{-1}}{1 + \left( \frac{z}{\ukk} \right)^2}.
\end{align}
Specifically, the vortex densities in the $(x^1,x^2)$- and 
the $(x^3,z)$-planes are evaluated as
\begin{align}
\tr \big[ F_{12} \big]
=& \ 
\frac{B_3}{3} + \frac{4}{N_c} \mu_B \Omega_3,
\notag \\
\tr \big[ F_{3z} \big]
=& \ - \frac{1}{f} \del_3 \eta \, \frac{2}{\pi} \frac{\ukk^{-1}}{1 + \left( \frac{z}{\ukk} \right)^2}
=
- \frac{1}{2} 
\Big(
\del_3 \phi_+ + \del_3 \phi_-
\Big)
\frac{2}{\pi} \frac{u_{\mathrm{kk}}^{-1}}{1 + \left( \frac{z}{\ukk} \right)^2}.
\end{align}
These induce source terms for D6-branes.
Indeed, $\tr [F_{12}]$ induces the constant source term in the worldvolume of the D8-branes:
\begin{align}
&
\mu_8 \lambda_s \int \! \tr \big[ F \big] \wedge C^{(7)} =
\mu_6 \mathbf{F} \varepsilon^{12 a_1 \cdots a_7} C^{(7)}_{a_1 \cdots a_7},
\notag \\
&
\mathbf{F} = \frac{1}{2\pi} \int \! dx^1 dx^2 \, 
\Big(
\frac{B_3}{3} + \frac{4}{N_c} \mu_B \Omega_3
\Big)
\quad 
(a_1, \ldots, a_7 = 0,3,z,6,7,8,9),
\end{align}
where $\mu_p$ is the D$p$-brane charge and we have shown only the relevant integrations.
Since the local density $\tr F_{12}$ is constant, the D6-brane charge is not localized, instead, it is uniformly distributed over the $(x^1,x^2)$-plane, consistent with the interpretation of dissolved D-branes.
On the other hand, $\tr [F_{3z}]$ generically induces a localized source term.
We have the couplings in the worldvolume of the D8-branes:
\begin{align}
&
\mu_8 \lambda_s \int \! 
\tr \big[ F \big] \wedge C^{(7)}
= 
-
\mu_6 
\,
\int_{-\infty}^{\infty}
\! d x^3 \,
\Big(
\del_3 \phi_+ + \del_3 \phi_-
\Big)
\,
\varepsilon^{3 z a_1 \cdots a_7} C^{(7)}_{a_1 \cdots a_7},
\notag \\
& 
(a_1, \ldots, a_7 = 0,1,2,6,7,8,9).
\end{align}
Since the CSL solutions \eqref{eq:phi_pm_amplitude} and \eqref{eq:phi_pm_kink} are localized in the $(x^3,z)$-plane, 
they form a periodic array of the D6 charge densities along $x^3$, localized in the holographic $z$-direction.
The vortex density in the $(x^3,z)$-plane is shown in Figure~\ref{fig:eta_CSL}.
However, the net charge $\int^{\infty}_{-\infty} \! dx^3 \, \Big( \del_3 \phi_+ + \del_3 \phi_- \Big)$ vanishes when $(\phi_+,\phi_-)$ is either (kink,anti-kink) or (anti-kink, kink).
See Table \ref{tb:D6kink}.
In this phase, only $\pi^0$ CSL survives.
On the other hand, for (kink,kink) and (anti-kink, anti-kink), the D6 charge does not cancel and the $\eta$ CSL remains.

\begin{table}[t]
\begin{center}
\begin{tabular}{c|c|c}
$(\phi_+,\phi_-)$ 
& (k,k), ($\bar{\mathrm{k}}$,$\bar{\mathrm{k}}$)
& (k,$\bar{\mathrm{k}}$), ($\bar{\mathrm{k}}$,k)
 \\
\hline
\hline
D6 charge 
& non-zero 
& zero 
\\
\hline
CSL 
& $\eta$
& $\pi^0$
\\
\hline
\end{tabular}
\end{center}
\caption{The net D6(03$z$6789) charge for the $\eta$-$\pi^0$ CSL. k for kink and $\bar{\mathrm{k}}$ for anti-kink. The relevant field is shown for CSL.}
\label{tb:D6kink}
\end{table}

\begin{figure}[t]
\begin{center}
\includegraphics[scale=0.5]{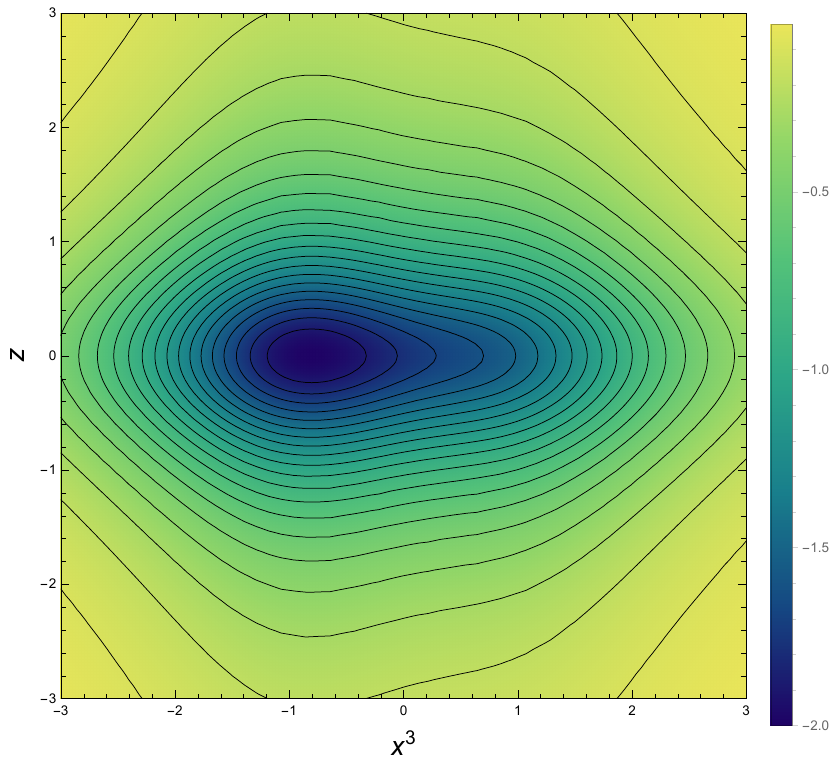}
\includegraphics[scale=0.5]{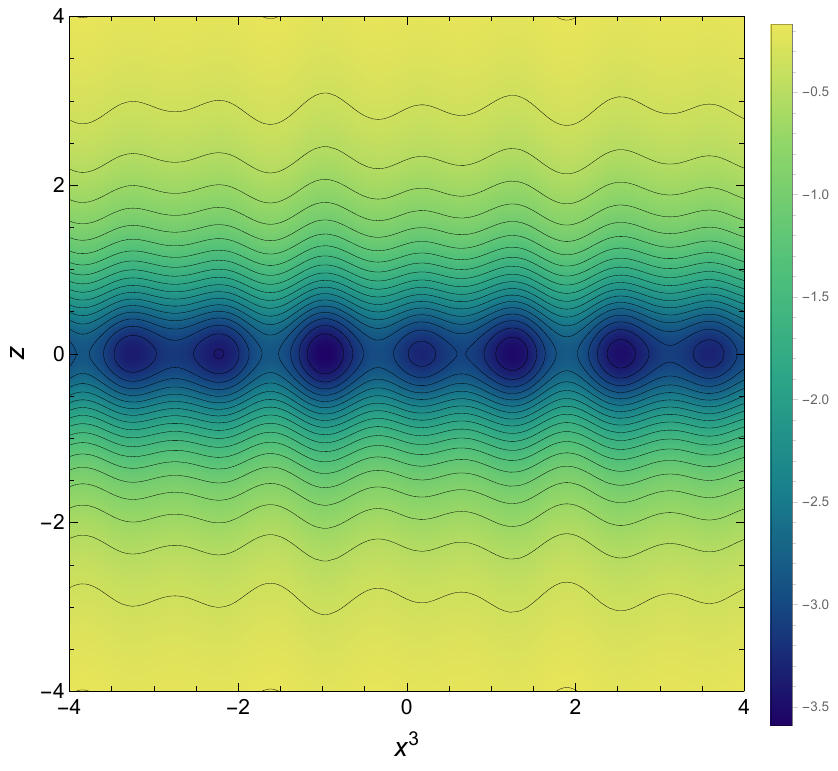}
\end{center}
\caption{
The instanton density $\tr [F_{12} F_{3z}]$ for $\eta$-$\pi^0$ CSL.
Left: 1-kinks for $\phi_+, \phi_-$ located at $x^+_0 = -1, \tilde x^+_0 = 1$.
Parameters $\mu_B, m, B_3, \Omega_3, \ukk, N_c$ are set to be 1.
}
\label{fig:eta-pi_CSL}
\end{figure}

The CSL together with the background magnetic field and rotation provides non-zero instanton density in the $(x^1,x^2,x^3,z)$-plane.
This is given by
\begin{align}
\tr \Big[ F_{12} F_{3z} \Big]
=& \ 
- 
\left\{
\left(
\frac{1}{3} B_3 + \frac{1}{N_c} \mu_B \Omega_3
\right)
\del_3 \phi_+
+
\left(
- \frac{1}{6} B_3 + \frac{1}{N_c} \mu_B \Omega_3
\right)
\del_3 \phi_-
\right\}
\frac{2}{\pi}
\frac{\ukk^{-1}}{1 + \left( \frac{z}{\ukk} \right)^2}.
\end{align}
This induces the coupling in the worldvolume of the D8-branes:
\begin{align}
&
\mu_8 \frac{\lambda_s^2}{2} \int \! 
\tr 
\big[
F \wedge F 
\big]
\wedge C^{(5)}
\notag \\
&=   
\mu_4 \, 
\int \! dx^1 dx^2 \,
\frac{(-2)}{(2\pi)^2}
\int_{-\infty}^{\infty} \! dx^3 \,
\left\{
\left(
\frac{1}{3} B_3 + \frac{1}{N_c} \mu_B \Omega_3
\right)
\del_3 \phi_+
+
\left(
- \frac{1}{6} B_3 + \frac{1}{N_c} \mu_B \Omega_3
\right)
\del_3 \phi_-
\right\}
\notag \\
& 
\hspace{1cm}
\times \varepsilon^{123z a_1 \cdots a_5}
\frac{1}{5!} C^{(5)}_{a_1 \cdots a_5}
\quad
(a_1, \ldots, a_5 = 0,6,7,8,9).
\end{align}
This indicates that the non-zero D4-brane charge is induced by the $\eta$-$\pi^0$ CSL in the presence of the background magnetic field and rotation.
The fact $\tr [ F_{12} F_{3z} ] \not= 0$ also implies that the $\eta$-$\pi^0$ CSL with $B$ and $\Omega$ carries the baryon charge \cite{Sakai:2004cn, Sakai:2005yt, Amano:2025iwi}.
The instanton density in the $(x^3,z)$-plane is shown in Figure~\ref{fig:eta-pi_CSL}.

We note that the brane interpretation discussed in this section is valid only within the leading order of $\Omega$, in which vorticity is encoded in a boundary $\mathrm{U}(1)_B$ gauge field.

\section{Bulk analysis}
\label{sec:bulk_analysis}
In the previous section, we have performed only the zero-mode analysis, namely, the dynamics relevant for $\pi^0$ and $\eta$ sectors.
In this section, we derive and solve the equations of motion for the action $S_{D8} = S_{\mathrm{DBI}} + S_{\mathrm{CS}}$ involving towers of massive vector mesons and the backreaction of general constant magnetic fields $B$ and vorticity $\Omega$.
In the absence of the external magnetic field $B$ and the angular velocity $\Omega$, the tower of the massive vector mesons does not contribute to the effective action of the pion and eta zero modes at the two-derivative order. 
However, when $B \not= 0$ or $\Omega \not= 0$, they induce the mixing of the vector mesons and zero modes through the WZW term.
We will show that the effects of the vector mesons result in the modification of the decay constant.
Since $S_{\mathrm{DBI}}$ is of order $\mathcal{O}(\lambda^1)$ and $S_{\mathrm{CS}}$ is of order  $\mathcal{O}(\lambda^0)$, we find that limiting our scope to $\mathcal{O}(\lambda^1)$  contributions reproduces ChPT \cite{Sakai:2004cn, Sakai:2005yt}.
Nevertheless, we can  also include $S_{\mathrm{CS}}$ terms to contribute to the equations of motion.
By doing so, we introduce a source term that is of order $\mathcal{O}(F^2)$. 
Finally, we analyze the  resulting condensed phase.

\paragraph{Action and equations of motion}

We use Sakai-Sugimoto units where
\begin{equation}
    \begin{aligned}
        (x^\mu, u, z) &\rightarrow u_{\textrm{KK}} (x^\mu, u, z), \\
        R &\rightarrow \sqrt[3]{\frac 94} u_{\textrm{KK}}. \\
    \end{aligned}
\end{equation}
In these units the mass scale is related to $\ukk$ as $\mkk = \ukk^{-1}$.
The antipodal DBI action can be written as \eqref{eq:dbi_antipodal_action}.
\begin{equation}
    \label{eq:dbi_antipodal_action}
    S_{\textrm{DBI}} = -\kappa \int d^5x \tr \left(
    \frac{1}{2u} F_{\mu\nu}F^{\mu\nu} + 
    u^3 F_{z\mu}F^{z\mu} 
    \right),
\end{equation}
where indices are raised by the flat 5D metric, $u^3 = 1 + z^2$, and 
$$\kappa = \frac{ 9 \pi^2 T \alpha' u_{\textrm{KK}}}{2 g_s} = \frac{\pi}{4}f^2\, ,$$
where $T$ is the tension of the D8-brane.
Nicely enough, this action is equivalent to a 5D Yang-Mills theory in curved 5D space metric \eqref{eq:dbi_antipodal_metric}.
\begin{equation}
    \label{eq:dbi_antipodal_metric}
    ds^2_{5\mathrm{D}} = u^2 ds^2_{4\mathrm{D}} + \frac{1}{u^2} dz^2.
\end{equation}
where  $ds^2_{4\mathrm{D}}$ is a 4D flat metric.
Together with the CS action, the total perturbative D8-brane action is \eqref{eq:dbi_total_action}.
\begin{equation}
    \label{eq:dbi_total_action}
    \begin{aligned}
        S_{D8} &= S_{\mathrm{DBI}} + S_{\mathrm{CS}}\\
               &= -\kappa \int \sqrt{-g} \left( \frac 12 \tr F^2 \right)  + \kappa\beta \int \omega_5(A)\\
    \end{aligned}
\end{equation}
where $\beta = N_c/(6\pi^3f^2)$, and gauge quantities are Hermitian.
Variations of $S_{D8}$ with respect to $A_\mu$ lead to five equations of motion.
\begin{equation}
    \frac{1}{\sqrt{-g}} \pd_B (\sqrt{-g} F^{BA}) = \frac{3}{4}\beta \sqrt{-g} F_{BC} F_{DE}\epsilon^{BCDEA}
\end{equation}
where we set $N_f = 2$ and restricted to generators in the Cartan subalgebra.
$\epsilon^{BCDEA}$ is the 5D volume form tensor density where $\epsilon^{BCDEA} = g \epsilon_{BCDEA}$.
Next we look at our specific ansatz.

\subsection{Massless case}
We first begin with the massless case $m = 0$.

\paragraph{Ansatz} 
To start, 
we first specify the external magnetic field at the boundary, allowing for arbitrary constant components,
\begin{align}
    F_{ij} (z \to \pm \infty) & = \varepsilon_{ijk} B_k, \\
    B_i (z \to \pm \infty) &= B_{i0} \tau^0 + B_{i3} \tau^3 = B_{Qi} Q + 2 \frac{\mu_B}{N_c} \Omega_i \tau^0,
\end{align}
where $B_i = B_{ia} \tau^a \, (a=0,3)$ are constants and $\tau^a$ are generators for the Cartan subalgebra of $\mathfrak u(2)$.
Note that the components $B_{i0}$, $B_{i3}$ are linear combinations corresponding to the magnetic field $B_{Qi}$ in $\mathrm{U}(1)_{\mathrm{em}}$ and the angular momentum together with the baryon chemical potential $\mu_B \Omega_i$ in $\mathrm{U}(1)_B$.
In the following, we leave the directions of $B_{Qi}$ and $\Omega_i$ arbitrary.
Also, to keep the presentation of equations and expression brief, we treat most bulk fields and quantities going forward as matrix valued quantities unless stated otherwise.
To be consistent with Section \ref{sec:chpt_csl_nf2} corresponding bulk gauge potential can be taken as
\begin{equation}
    A = A_t(z) dt + A_i(z) dx^i + b_i(z) \varepsilon_{ijk} x^j dx^k, 
\end{equation}
where $b_i(z)$ reduces to the constant boundary field strengths,
\begin{equation}
    \label{eq:bulk_b_boundary_conditions}
    b_i(z = \pm\infty) = B_{i0} \tau^0 + B_{i3} \tau^3,
\end{equation}
and we impose $b_i = B_i$ to solve two equations of motion and \eqref{eq:bulk_b_boundary_conditions}.
Boundary values for the other field components are as follows:
\begin{equation}
    \label{eq:boundary_conditions}
    \begin{aligned}
        A_t(z=\pm\infty) & = \frac{\mu_B}{N_c} \tau^0 \equiv \mu, \\
        A_i(z=\pm\infty) &= \mp\frac 12\pd_i\phi_0\tau^0  \mp \frac 12 \pd_i\phi_3 \tau^3 
        = \mp\frac 12\pd_i\left(\frac{\eta}{f}\right) \tau^0 \mp \frac 12\pd_i\left(\frac{\pi^0}{f}\right) \tau^3.
    \end{aligned}
\end{equation}
The factor of $1/2$ is needed for consistency with the $A_z = 0$ gauge as explained in \cite[pg. 14]{Amano:2025iwi}.
Here $\mu_B$ is a constant identified with the baryon chemical potential and 
$\phi_a = \phi_a (x) \, (a=0,3)$ are linear functions of $x^{\mu}$ at the boundary \cite{Amano:2025iwi}.
In addition, to eliminate the gauge redundancy, we choose the holographic gauge where $A_z = 0$.
Evaluating the equations of motion, we obtain two coupled sets of ODEs for $F_{zt}$ and $F_{zi}$,
\begin{equation}
    \label{eq:massless_coupled_ODEs}
    \begin{aligned}
        \pd_z \left(K(z) F_{z t}\right) &= - \mathcal B_i \delta^{ij} F_{zj},\\
        \pd_z \left(K(z) F_{zi}\right) &= - \mathcal B_i F_{zt},\\
    \end{aligned}
\end{equation}
where $\mathcal B_{03} = 6\beta B_{03} \equiv N_c/(f^2 \pi^3) B_{03}$, and $\mathcal B\equiv \mathcal B_{ia} \tau^a dx^i$.
The derivation of the bulk solutions satisfying \eqref{eq:massless_coupled_ODEs} subject to the boundary conditions \eqref{eq:boundary_conditions} is carried out in Appendix~\ref{app:massless_bulk_solutions}.
We work in the $\tau^\pm := \frac12(\tau^0 \pm \tau^3)$ basis with $\mathcal{B}_{i\pm} = \mathcal{B}_{i0} \pm \mathcal{B}_{i3}$ and the transverse projector $\mathcal{P}_{ij\pm} := \delta_{ij} - \hat{\mathcal{B}}_{i\pm}\hat{\mathcal{B}}_{j\pm}$ where the equations of motion decouple.
This basis applies to other $\mathrm{SU}(2)$-valued variables such as $\mu_\pm=\mu_0 \pm \mu_3$ and $\phi_\pm=\phi_0 \pm \phi_3$.
The key result is that the on-shell bulk fields read
\begin{equation}
    \label{eq:massless_bulk_solution}
    \begin{aligned}
        F_{\xi i\pm} &= -\left( \frac{1}{2}|\mathcal B_\pm|\frac{\cosh(|\mathcal{B}_\pm|\xi)}{\sinh(|\mathcal{B}_\pm|\pi/2)} \hat{\mathcal B }_{i\pm} \hat{\mathcal B}_{j\pm} + \frac{1}{\pi} \mathcal{P}_{ij\pm} \right)  \pd_j\phi_\pm, \\
        A_{i\pm} &= - \left(\frac{1}{2}\frac{\sinh(|\mathcal{B}_\pm|\xi)}{\sinh(|\mathcal{B}_\pm|\pi/2)} \hat{\mathcal B }_{i\pm} \hat{\mathcal B}_{j\pm} + \frac{1}{\pi} \mathcal{P}_{ij\pm} \xi \right)  \pd_j\phi_\pm, \\
        F_{\xi t\pm} &= \frac{1}{2}|\mathcal B_\pm|\frac{\sinh(|\mathcal{B}_\pm|\xi)}{\sinh(|\mathcal{B}_\pm|\pi/2)} \hat{\mathcal{B}}_{j\pm}\pd_j\phi_\pm,  \\
        A_{t\pm} &= \frac{1}{2}\frac{\cosh(|\mathcal{B}_\pm|\xi) - \cosh(|\mathcal{B}_\pm|\pi/2)}{\sinh(|\mathcal{B}_\pm|\pi/2)} \hat{\mathcal{B}}_{j\pm}\pd_j\phi_\pm + \mu_\pm, \\
    \end{aligned}
\end{equation}
where $\xi = \arctan z$ and $\hat{\mathcal{B}}_\pm$ is the unit vector of $\mathcal{B}_\pm$.
$|\mathcal{B}_\pm| := \sqrt{\mathcal{B}_{i\pm}\mathcal{B}_{i\pm}}$ is the magnitude of $\mathcal{B}_{\pm}$.

\paragraph{On shell action}
Compared to the single $B$ component case, we can add a few more terms to make the  $S_{\mathrm{DBI}}$ correct for the multi component case.
In addition, we also neglect  the extra divergent square components of $B_{1,2}$ since they are not dynamical.
The on-shell action and the Hamiltonian  for the DBI part are given by
\begin{equation}
  \begin{aligned}
      S_\mathrm{DBI} &= \kappa \int d\xi d^4x \tr \left(-F_{\xi i} F_{\xi i} + F_{\xi t}^2 \right), \\
      H_\mathrm{DBI} &= \kappa \int d\xi d^3x \tr \left( F_{\xi i} F_{\xi i} + F_{\xi t}^2 \right). \\
  \end{aligned}
\end{equation}
The CS part will need some extra care however.
We find they are given by
\begin{equation}
   \begin{aligned}
      S_\mathrm{CS} &= \frac 23 \kappa \int d^4x \left(-\int d\xi \tr\left((A_t - \mu) \mathcal B_i F_{\xi i} \right) + \frac 32\tr\left(\mu \mathcal B_i \pd_i\phi \right) \right),\\
      H_\mathrm{CS} &= \frac\pi6 f^2 \int d^3x \left(\int d\xi \tr\left((A_t - \mu) \mathcal B_i F_{\xi i} \right) - \frac 32\tr\left(\mu \mathcal B_i \pd_i\phi \right) \right),\\
   \end{aligned} 
\end{equation}
where $\mu = \mu_B/N_c \tau^0$.
We also had to account for how $S_\mathrm{CS}$ transforms anomalously.

Let us now compute them using the on-shell fields.
Evaluating the $\xi$-integration, we find
\begin{equation}
    \begin{aligned}
        H_\mathrm{DBI} &= \frac\pi{16} f^2 V_3 \sum_\pm \left(
            \frac{|\mathcal B_\pm| \sinh(|\mathcal B_\pm| \pi)}{\sinh^2(|\mathcal{B}_\pm|\pi/2)}  (\hat{\mathcal B}_{j\pm} \pd_j\phi_\pm)^2 +
            \frac{4}{\pi} \left(\mathcal{P}_{ij\pm}\pd_j\phi_\pm\right)^2
        \right), \\
    \end{aligned}
\end{equation}
\begin{equation}
    \begin{aligned}
        H_\mathrm{CS} &= -\frac\pi6 f^2 V_3 \sum_\pm \left(\frac 18\frac{\pi |\mathcal B_\pm|^2 - |\mathcal B_\pm|\sinh(\pi |\mathcal B_\pm|)}{\sinh^2(|\mathcal{B}_\pm|\pi/2)} \left( \hat{\mathcal B}_{j\pm} \pd_j\phi_\pm \right)^2 + \frac 32 \mu_\pm \mathcal B_{i\pm} \pd_i\phi_\pm \right), \\
    \end{aligned}
\end{equation}
where we have used the fact that $\del_j \phi_{\pm}$ are constants.
Altogether we can write out the Hamiltonian  in four dimensions as
\begin{equation}
    H_\mathrm{4D} = \frac 14 f^2 V_3 \sum_\pm \left(
        \pd_i \phi_\pm \left(\frac \pi{12} \frac {4 |\mathcal B_\pm| \sinh(|\mathcal B_\pm| \pi) -\pi |\mathcal B_\pm|^2}{\sinh^2(|\mathcal{B}_\pm|\pi/2)} \hat{\mathcal B}_{i\pm} \hat{\mathcal B}_{j\pm} + \mathcal{P}_{ij\pm} \right)\pd_j \phi_\pm - \frac\pi{12} \mu_\pm \mathcal B_{i\pm} \pd_i\phi_\pm \right).
\end{equation}

We can write our Hamiltonian as
\begin{equation}
    \label{eq:bulk_chiral_4D_hamiltonian}
    H_{4D} = V_3 \left(\sum_{\pm} \frac 14 \pd_i\phi_\pm\tilde f^2_{ij\pm}\pd_j\phi_\pm - \frac\pi{4} f^2 \mu_\pm \mathcal B_{i\pm} \pd_i\phi_\pm\right),
\end{equation}
and
\begin{equation}
    \tilde f^2_{ij\pm} = f^2 \frac \pi{12} \frac {4 |\mathcal B_\pm| \sinh(|\mathcal B_\pm| \pi) -\pi |\mathcal B_\pm|^2}{\sinh^2(|\mathcal{B}_\pm|\pi/2)} \hat{\mathcal B}_{i\pm} \hat{\mathcal B}_{j\pm} + f^2 \mathcal{P}_{ij\pm}.
\end{equation}
For small $\mathcal B_\pm$ we have sub-leading quadratic order as
\begin{equation}
    \tilde f^2_{ij\pm} \sim f^2 \delta_{ij} + \frac{5 f^2\pi^2 }{36} \mathcal B_{i\pm} \mathcal B_{j\pm}
    \qquad (\text{small-}\mathcal{B}_{\pm}).
\end{equation}
For large $\mathcal B_\pm$, we have linear behavior as
\begin{equation}
    \tilde f^2_{ij\pm} \sim f^2 \frac {2\pi |\mathcal B_\pm|}{3} \hat{\mathcal B}_{i\pm} \hat{\mathcal B}_{j\pm}
    \qquad (\text{large-}\mathcal{B}_{\pm}).
\end{equation}

Solving for the two $\pd\phi_\pm$ that minimize the energy, we can see that ground states' gradients are parallel to $\mathcal B$.
The ground states can be written as
\begin{equation}
    \label{eq:wss_min_del_phi}
    \pd_i\phi_\pm = 
    \frac{6\mu_\pm \sinh^2(|\mathcal B_\pm|\pi/2)}
    {4 \sinh(|\mathcal B_\pm|\pi) - \pi |\mathcal B_\pm|} \hat{\mathcal B}_{i\pm}.
\end{equation}
Note for general $\mathcal B$, \eqref{eq:wss_min_del_phi} is parallel to $\mathcal B$ while perpendicular components vanish.
Adapting our coordinates such that the $x^3$ direction aligns with $\mathcal B$, one can recover our previous result in \cite{Amano:2025iwi}.
In the large magnetic field limit, we have
\begin{equation}
    \pd_i\phi_\pm \sim \frac {3\mu_\pm}4  \hat{\mathcal{B}}_{i\pm} \qquad (\text{large-}\mathcal{B}_{\pm}).
\end{equation}
The gradient of the scalar field 
$\pd\phi_\pm$ in the large $B$ limit lies on a 2-sphere where the radius is $\frac {3\mu_\pm}4$.
For weaker magnetic fields, ground states can be shown to lie inside this sphere, parameterized by $\mathcal B_\pm$.
Connecting back with ChPT, we can expand for small fields which is in agreement with ChPT with leading order $\mathcal O(\mathcal B_\pm)$,
\begin{equation}
   \pd_i\phi_\pm \sim  \mu_\pm \frac{\pi}{2} \mathcal B_{i\pm}
    \qquad (\text{small-}\mathcal{B}_{\pm}).
\end{equation}

The minimum energy density can be written as
\begin{align}
    \frac{H}{V_3} \Big|_\mathrm{min} &= 
    -\frac{3\pi f^2}{8} \sum_\pm \frac{ \mu_\pm^2|\mathcal B_\pm| (\cosh (\pi  |\mathcal B_\pm|)-1)}{4 \sinh
    (\pi  |\mathcal B_\pm|)-\pi  |\mathcal B_\pm|}
    =
    -\frac 12 \sum_\pm \mu_\pm n_\pm, \\
    n_\pm &= \frac{3\pi f^2}{4}\frac{ \mu_\pm|\mathcal B_\pm| (\cosh (\pi  |\mathcal B_\pm|)-1)}{4 \sinh (\pi  |\mathcal B_\pm|)-\pi  |\mathcal B_\pm|},
\end{align}
where $n_\pm$ are the number densities with respect to $\pm$ basis.
For small magnetic/vortical fields, the energy density $\left.\frac{H}{V_3}\right\vert_\mathrm{min}$ is evaluated as 
\begin{equation}
    \label{eq:ss_smallB_min_energy}
    \left.\frac{H}{V_3}\right\vert_\mathrm{min} \approx 
    -\sum_\pm\left( \frac{\pi f \mu_\pm}{4} |\mathcal B_{\pm}|\right)^2
    \qquad (\text{small-}B_{\pm}).
\end{equation}
In the large-$\mathcal{B}_{\pm}$ limit the energy depends linearly on the fields independent of $f$,
\begin{equation}
    \label{eq:ss_largeB_min_energy}
    \left.\frac{H}{V_3}\right\vert_\mathrm{min} \approx 
    -\frac{3f^2\pi}{32}\sum_\pm \mu_\pm^2 |\mathcal B_{\pm}|
    \qquad (\text{large-}B_{\pm}).
\end{equation}
One can see that from small to large magnetic fields, the magnetization changes from being linear with respect to the magnetic field to being constant.
This seems to indicate that there is a saturation effect.
At $|\mathcal B_\pm| \approx 0.57168$, magnetization maximizes and for larger magnetic field, the magnetization decreases with respect to increasing $|\mathcal B_\pm|$.
Another way to see this is in terms of angular momentum susceptibility.
For small chiral fields, these susceptibilities are linear.
But for strong field strengths, these susceptibilities asymptote to a constant value with respect to $\mathcal B$.

\subsection{Massive case}

We next introduce a mass term for the meson fields.
Using the units $R$ and $u_{\mathrm{KK}}$ in \cite{Sakai:2004cn}, we can write down the equations of motion for the field strengths, similar to \cite{Amano:2025iwi}:
\begin{equation}
  \begin{aligned}
      \pd_3 (u^3 F_{z3}) &= -\mathcal BF_{3t} - m^2 \sum_\pm  \sin(\phi_\pm) \tau^\pm, \\
      u^{-1} \pd_3 F_{3t} +  \pd_z \left(u^3 F_{zt}\right) &= -\mathcal B F_{z3}, \\
      \pd_z \left(u^3 F_{z3}\right) &= -\mathcal B F_{zt},\\
  \end{aligned}
\end{equation}
where we have assumed that the magnetic field and the angular momentum are introduced along the $x^3$ direction, $B_3$, $\Omega_3$ and  $\mathcal{B} = \mathcal{B}_{3a} \tau^a$.
Here we have assumed that the bulk magnetic fields are constant.
Further, the boundary value of $A_3 = -\frac{1}{2}\pd_3 \phi$ is now assumed to be a function of $x^3$.
The mass term can be derived (see the appendix of \cite{Amano:2025iwi}).
Writing in terms of vector potentials, the equations of motion can be rewritten.
\begin{equation}
  \begin{aligned}
      \pd_3 (u^3 \pd_zA_3) &= -\mathcal B\pd_3A_t - m^2 \sum_\pm  \sin(\phi_\pm) \tau^\pm 
, \\
      u^{-1} \pd_3^2 A_t +  \pd_z \left(u^3 \pd_zA_t\right) &= -\mathcal B \pd_zA_3, \\
      \pd_z \left(u^3 \pd_zA_3\right) &= -\mathcal B \pd_zA_t. \\
  \end{aligned}
\end{equation}

Observe that the first and third equations imply $\pd_\xi A_3 + \mathcal B A_t = C (x^3)$ where $\pd_3 C = -m^2 \sum_\pm  \sin(\phi_\pm) \tau^\pm$.
The value $C$ can be determined by using the boundary conditions of $A_3$ and $A_t$.
A consistent value of $C$ is $C = -\pd_3 \phi/\pi + Q$ where
$\pd_3^2\phi /\pi= m^2 \sum_\pm  \sin(\phi_\pm) \tau^\pm$.
Note, that $Q$ just parameterizes the residual gauge freedom of $A_t$ where $A\rightarrow A_t+\mathrm{const}$ also solves the bulk equations of motion.
Note that the $\xi = \arctan(z)$ coordinate is used for this section.
The second equation of motion can now be written in terms of $F_{\xi3}$.
\begin{equation}
    \left( u^2 \pd_3^2 +  \pd_\xi^2 - \mathcal B^2 \right) F_{\xi3} = -\frac 1\pi u^2 \pd_3^3\phi \Longleftrightarrow
    \left( u^2 \pd_3^2 +  \pd_\xi^2 - \mathcal B_\pm^2 \right) F_{\xi3\pm} = -\frac 1\pi u^2 \pd_{3}^3\phi_\pm
\end{equation}
where $\phi = \phi_+ \tau^+ + \phi_- \tau^-$ and the equations were written in the $\tau^\pm$ basis.
Our boundary variables, $\phi_{\pm}$, thus obey the following second order non-linear ODE:
\begin{equation}
    \pd_3^2 \phi_\pm = m^{2} \sin(\phi_\pm).
\end{equation}
The solutions are \eqref{eq:chtp_csl_eom}.
Now let's write out the equation of motion  for $F_{\xi3}$:
\begin{equation}
    \left( u^2 \pd_3^2 + \pd_\xi^2 - \mathcal B_\pm^2 \right) F_{\xi3\pm}  = -\frac1\pi u^2 \pd_3^3\phi_\pm.
\end{equation}

This ODE is separable with a non-trivial inhomogeneous term.
Note that we are taking the positive branch of $\phi$.
So it is readily solvable with numerical methods.
Noting that the inhomogeneous term is even and periodic, we can decompose our $x^3$ direction in terms of cosine functions.
The Fourier transformation of the third derivative of the 
Jacobi amplitude function is prima facie non-trivial, but we can relate its Fourier coefficients to its first derivative.
Luckily, we can write down $\dn$'s Fourier series explicitly \cite{Olver:2010fr}.
\begin{equation}
    \begin{aligned}
        \dn(x, k) &= a_0 + \sum_{n=1}^\infty a_{n} \cos(2n\zeta) 
        = \frac{\pi}{2\mathcal K(k)} + \sum_{n=1}^\infty \frac{2\pi}{\mathcal K}\frac{q^n}{1+q^{2n}} \cos(2n\zeta)\\
        \zeta &= \frac{\pi x}{2\mathcal K(k)}, \quad
        q = \exp(-\pi \mathcal K'(k)/\mathcal K(k)), \quad
        \mathcal K' := \mathcal K\left(\sqrt{ 1-k^2 }\right). \\
    \end{aligned}
\end{equation}
Note, the implicit definition of $a_n$ as a function of $k$.
We can write the ODE per boundary harmonic in the $\theta_\pm := m\pi x^3/(k_\pm \mathcal K(k_\pm))$ coordinate such that the period is $2\pi$ valid if $0\leq k\leq 1$.
We also decompose $F_{\xi3\pm} = \sum_{n=0} \Xi_{n\pm}(\xi) \cos(n\theta_\pm)$ and find
\begin{equation}
    \pd_\xi^2 \Xi_{n\pm} 
    - u^2 \left(\frac{n m\pi}{k_\pm \mathcal K(k_\pm)}\right)^2 \Xi_{n\pm}
    - \mathcal B_\pm^2 \Xi_{n\pm}
    = u^2 \frac 1\pi \frac{2m}{k_\pm} \left(\frac{n m\pi}{k_\pm \mathcal K(k_\pm)}\right)^2 a_n(k_\pm)
    \\
\end{equation}
where $u^2 = \sec^{4/3}(\xi)$.
At the boundaries, the equation of motion for $n>0$ harmonics reduces to the following condition:
\begin{equation}
    \label{eq:fxi3_ng0_nearboundary_ode}
    \Xi_{n\pm}\left(\xi\sim\pm\frac\pi2\right) \sim -\frac 1\pi \frac{2m}{k_\pm}  a_{n}(k_\pm) \equiv -\frac 1\pi (\pd_3\phi_\pm)_{n}.
\end{equation}
Here $(\pd_3\phi_\pm)_n$ is the n\textsuperscript{th} harmonic of $\pd_3\phi$.
In general we will denote the n\textsuperscript{th} harmonic of $x$ as $(x)_n$.

\paragraph{$n > 0$ case}
For the $\xi$ direction, we can use Chebyshev polynomials for decomposition.
We note that $\Xi_{n\pm}$ for $n > 0$ must be such that $A_{3\pm}$'s boundary conditions are satisfied.
Due to the parity of the $\Xi_{n\pm}$ equations of motion and the boundary conditions on $A_3$, we can infer that $\Xi_{n\pm}$ are even functions.
There are two solutions to the $\Xi_{n\pm}$ equations of motion,
$\Xi_{n\pm} = \zeta_n \Xi_{nh\pm} + \Xi_{np\pm}$.
Per harmonic, the integrated value of $\Xi_{n\pm}$ must be $-\frac{1}{2}\pd_3\phi_\pm \supset -\frac{m}{k_\pm} a_n(k_\pm) \cos(n\theta_\pm)$ at the boundaries:
\begin{equation}
    \label{eq:massive_pion_zetan_bc}
    \zeta_{n\pm}  = \frac{-\frac{m}{k_\pm} a_n(k_\pm) - \int_0^{\pi/2}\Xi_{np\pm}d\xi} {\int_0^{\pi/2}\Xi_{nh\pm}d\xi}.
\end{equation}
Recall that we set $A_{3\pm}(z=0) = 0$, so that it is an odd function 
with respect to $z$.
Let's now consider the case, $n = 0$.

\paragraph{$n = 0$ case}
The $\xi$ dependent equation in this case is solvable:
\begin{equation}
    \pd_\xi^2 \Xi_{0\pm} - \mathcal B_\pm^2 \Xi_{0\pm} = 0. \\
\end{equation}
The solution is
\begin{equation}
    \begin{aligned}
        \Xi_{0h\pm} &= \cosh(\mathcal B_\pm \xi),\\
        \Xi_{0p\pm} &= 0.
    \end{aligned}
\end{equation}
With \eqref{eq:massive_pion_zetan_bc}, we can solve for $\zeta_{0\pm}$,
\begin{equation}
    \zeta_{0\pm}  = -\frac{m a_0(k_\pm)}{k_\pm} \frac{\mathcal B_\pm} {\sinh(\mathcal B_\pm \pi/2 )}.
\end{equation}
Plugging this in, we get a full solve for $\Xi_0$:
\begin{equation}
    \Xi_{0\pm} = -\frac{m a_0(k_\pm)}{k_\pm} \frac{\mathcal B_\pm \cosh(\mathcal B_\pm \xi)} {\sinh(\mathcal B_\pm \pi/2 )}.
\end{equation}

\paragraph{$A_{t\pm}$ Boundary conditions}

According to the equations of motion, we can solve for $A_{t\pm}$ in terms of $A_{3\pm}$/$F_{\xi3\pm}$:
\begin{equation}
    A_{t\pm} = - \frac 1{\mathcal B_\pm} \left(F_{\xi3\pm} - \left(F_{\xi3\pm}\right)_0\vert_{\xi = \pi/2} 
          + \frac{1}{\pi}\left( \pd_3\phi_\pm - (\pd_3\phi_\pm)_0 \right)\right) + \mu_\pm.
\end{equation}
This equality implies that $Q_\pm = (F_{\xi3\pm})_0\vert_{\xi = \pi/2} + \frac{1}{\pi}(\pd_3\phi_\pm)_0 - \mathcal B_\pm \mu_\pm$. Let's write out the $A_t$ per harmonic.
We can use \eqref{eq:fxi3_ng0_nearboundary_ode} to simplify the expression with some cancellations:
\begin{equation}
    A_t = - \mathcal B^{-1} \left(F_{\xi3} - F_{\xi3}\vert_{\xi = \pi/2} 
          \right) + \mu.
\end{equation}

\paragraph{Form of solution}

With a numerical solution for $F_{\xi3}$, we can subsequently analytically find the solutions of the other fields:
\begin{equation} 
  \begin{aligned}
      A_{t\pm} &= -\frac 1 {\mathcal B_\pm} \left(F_{\xi3\pm} - F_{\xi3\pm}\vert_{\xi = \pi/2}\right) + \mu_\pm,\\
      A_{3\pm} &= \int_0^\xi F_{\xi3\pm} d\xi, \\
      F_{\xi t\pm} &= \pd_\xi A_{t\pm} \equiv  -\frac 1 {\mathcal B_\pm} \pd_\xi F_{\xi3\pm},\\
      F_{3t\pm} &= \pd_3 A_{t\pm}.\\
  \end{aligned}
\end{equation}

\paragraph{On shell action}
Now that the equations of motion have been solved semi-analytically, we can partially evaluate and analytically study the on-shell action.
For an approximate value of $k$, we assume that the energy admits a global minimum.
Again, we disregard the $B^2$ term from the $S_\mathrm{DBI}$ since it does not depend on $k$.
$\kappa$ and $\beta$ are chosen to be consistent with the rest of the section:
\begin{equation}
   \begin{aligned}
      S_\mathrm{DBI} &= \kappa  V_3 \int dx^3 d\xi \tr \left( u^2 F_{3t}^2 - F_{\xi 3}^2 + F_{\xi t}^2 \right), \\
   \end{aligned} 
\end{equation}
\begin{equation}
   \begin{aligned}
      S_\mathrm{CS} &= \kappa V_3\left( -\int dx^3 d\xi \tr\left( \frac 23 \mathcal B (A_{t\pm} - \mu_\pm)  F_{\xi 3} \right) + \int dx^3 \tr\left( \mathcal B \pd_3\phi \mu \right) \right),\\
   \end{aligned} 
\end{equation}
\begin{equation}
    \begin{aligned}
        S_\mathrm{mass} &= \frac{\kappa m^2}{\pi} \int d^4x \tr (U + U^\dagger - 2),\\
        \frac{S_\mathrm{mass}}{\kappa V_2} &=-\sum_\pm T_\pm\frac{4 m^2 \left(\left(k_\pm^2-1\right) \mathcal K(k_\pm)+E(k_\pm)\right)}{\pi  k_\pm^2 \mathcal K(k_\pm)}\,.\\
    \end{aligned}
\end{equation}

In terms of our fields, the total density can be written as
\begin{equation}
    \begin{aligned}
        \frac{H}{\kappa V_2} &=\int dx^3 d\xi \tr \left(u^2 F_{3t}^2 + F_{\xi 3}^2 + F_{\xi t}^2 \right) \\
                             &+\int dx^3\left(\int d\xi \tr\left(  \frac 23  \mathcal B\tilde A_t F_{\xi 3} \right) - \tr\left( \mathcal B \mu \pd_3\phi\right) \right)\\
                             &+\sum_\pm T_\pm \frac{4 m^2 \left(\left(k_\pm^2-1\right) 
                             \mathcal K(k_\pm)+E(k_\pm)\right)}{\pi  k_\pm^2 \mathcal K(k_\pm)}.
    \end{aligned}
\end{equation}
Here $T_\pm$ are the periods of each $\pm$ sector.
Note that we do not include the $\mathcal B^2$ term here since it ends up diverging when integrated to the boundary, so we treat $\mathcal B$ as fixed external fields.
One might notice that the anomalous term and mass term are equivalent to the analogous terms in ChPT.
To this end, we can identify the remaining terms as the ``kinetic energy'' density, $H_{KE}/V_3$, which we identify with the ``$\frac{f^2}{4}\int dx^3 \sum_\pm (\del_3 \phi_\pm)^2$'' term in ChPT:
\begin{equation}
    \frac{H_{KE}}{\kappa V_2} =\int dx^3 d\xi \tr \left(u^2 F_{3t}^2 + F_{\xi 3}^2 + F_{\xi t}^2 + \frac 23  \mathcal B (A_{t\pm} - \mu_\pm) F_{\xi 3}\right).
\end{equation}

\paragraph{Numerical evaluation}
We can numerically evaluate by a harmonic decomposition along the $x^3$ direction as
\begin{equation}
   \begin{aligned}
      \frac{H_\mathrm{DBI}}{\kappa V_2} &= \sum_\pm T_\pm \int d\xi \sum_{n=0} \frac{(1 + \delta_n^0)}{2} \left( u^2 (F_{3t\pm})_n^2 + (F_{\xi 3\pm})_n^2 + (F_{\xi t\pm})_n^2 \right), \\
      \frac{H_\mathrm{CS}}{\kappa V_2} &= \sum_\pm T_\pm \left(\int d\xi \sum_{n=0} \frac{(1 + \delta_n^0)}{2} \left( \frac 23 \mathcal B_\pm ((A_{t\pm})_n - \mu_\pm \delta_n^0) (F_{\xi 3\pm})_n \right) - \frac{m}{2k_\pm \mathcal K(k_\pm)}\left| \left(2\pi \mu_\pm \mathcal B_\pm\right) \right| \right),\\
      \frac{H_\mathrm{mass}}{\kappa V_2} &= \sum_\pm T_\pm \frac{4 m^2 \left(\left(k_\pm^2-1\right) 
                             \mathcal K(k_\pm)+E(k_\pm)\right)}{\pi  k_\pm^2 \mathcal K(k_\pm)}.
   \end{aligned} 
\end{equation}
Just like the ChPT case, each $\pm$ sector can be minimized seperately to determine the stability of the bulk CSL$_\pm$ phases respectively.
In the actual calculation, we divide both $\pm$ energies by $T_\pm$.

\subsection{Phase diagrams}

\begin{figure*}[t]
  \centering
  \includegraphics[width=0.33\textwidth]{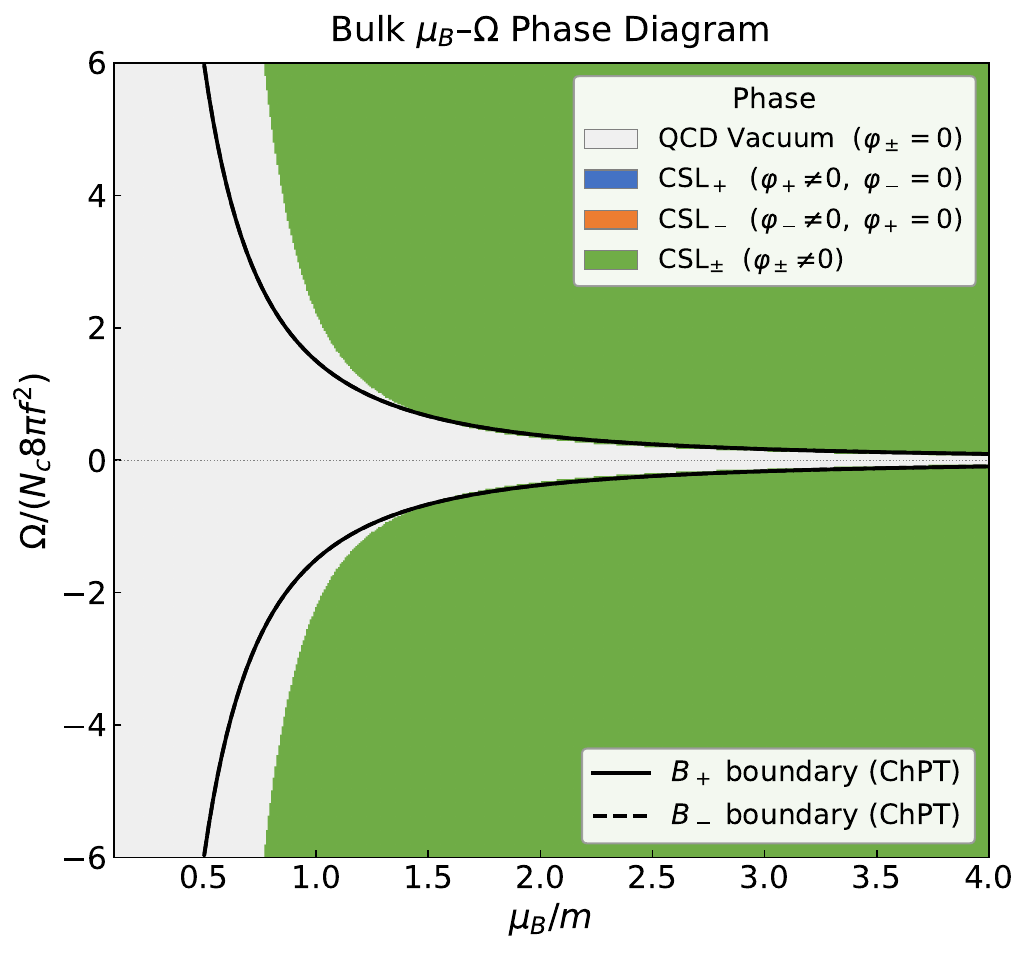}%
  \includegraphics[width=0.33\textwidth]{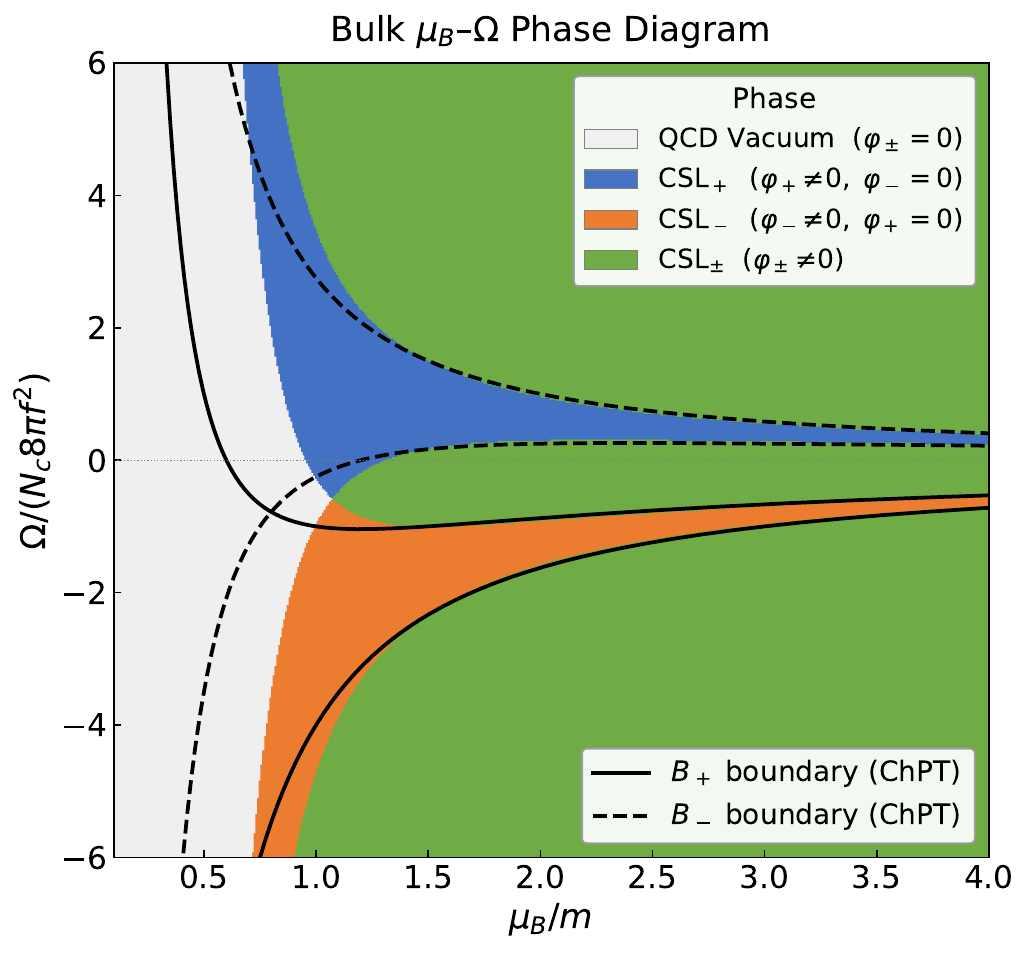}%
  \includegraphics[width=0.33\textwidth]{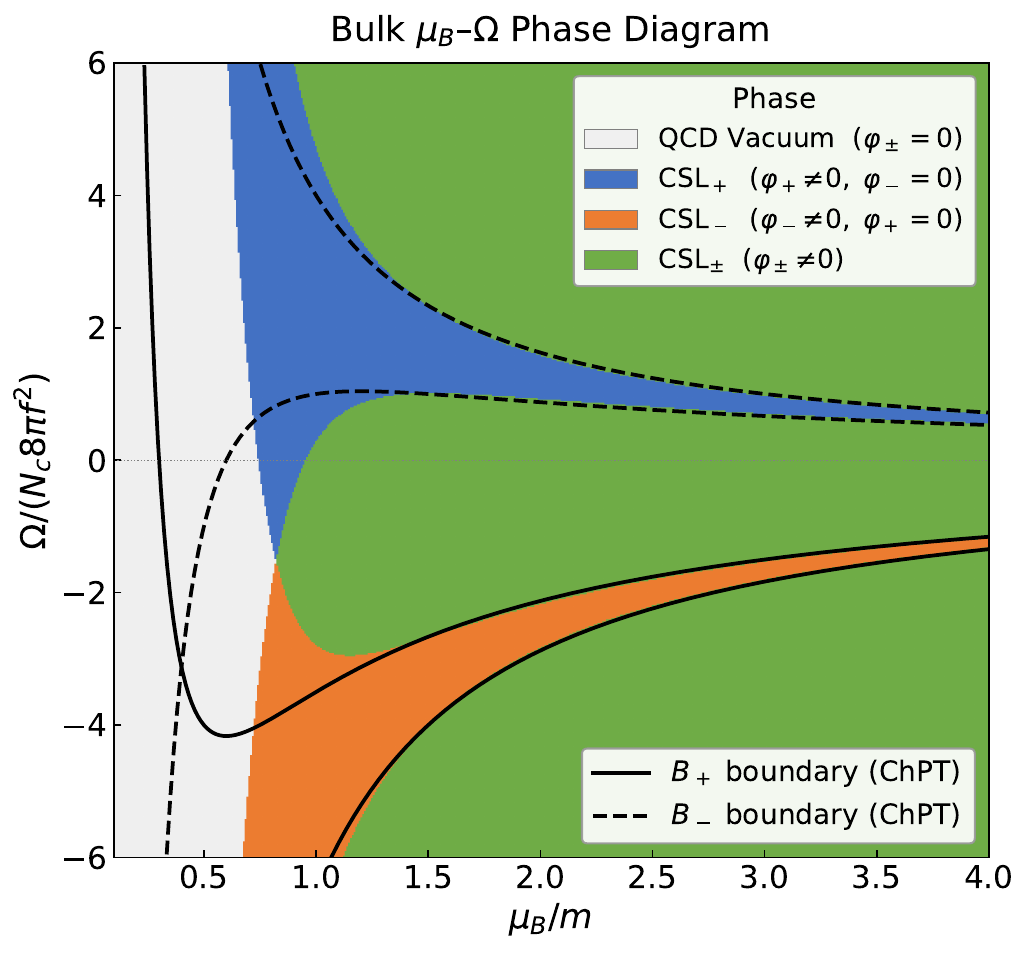}
  \caption{Bulk phase diagrams in the $\mu_B$--$\Omega$ plane, with axes shown
    as the dimensionless ratios $\mu_B/m$ and $\Omega/(N_c\,8\pi f^2)$.
    Left: $B/(8\pi f^2) = 0$, where $B_+ = B_-$ and the two CSL boundaries are degenerate,
    so only the QCD vacuum and $\mathrm{CSL}_\pm$ phases survive.
    Middle: $B/(8\pi f^2) = 2.5$ and Right: $B/(8\pi f^2) = 5.0$, where the degeneracy is
    lifted and all four phases appear, with the $\mathrm{CSL}_-$ band widening as $B$ increases.
    Solid (dashed) curves are the ChPT $B_+$ ($B_-$) boundaries, given by
    $(\mu_B/m)\,|B_\pm|/(8\pi f^2) = 1$.}
  \label{fig:bulk_phase_mu_Omega}
\end{figure*}

\begin{figure*}[h]
  \centering
  \includegraphics[width=0.33\textwidth]{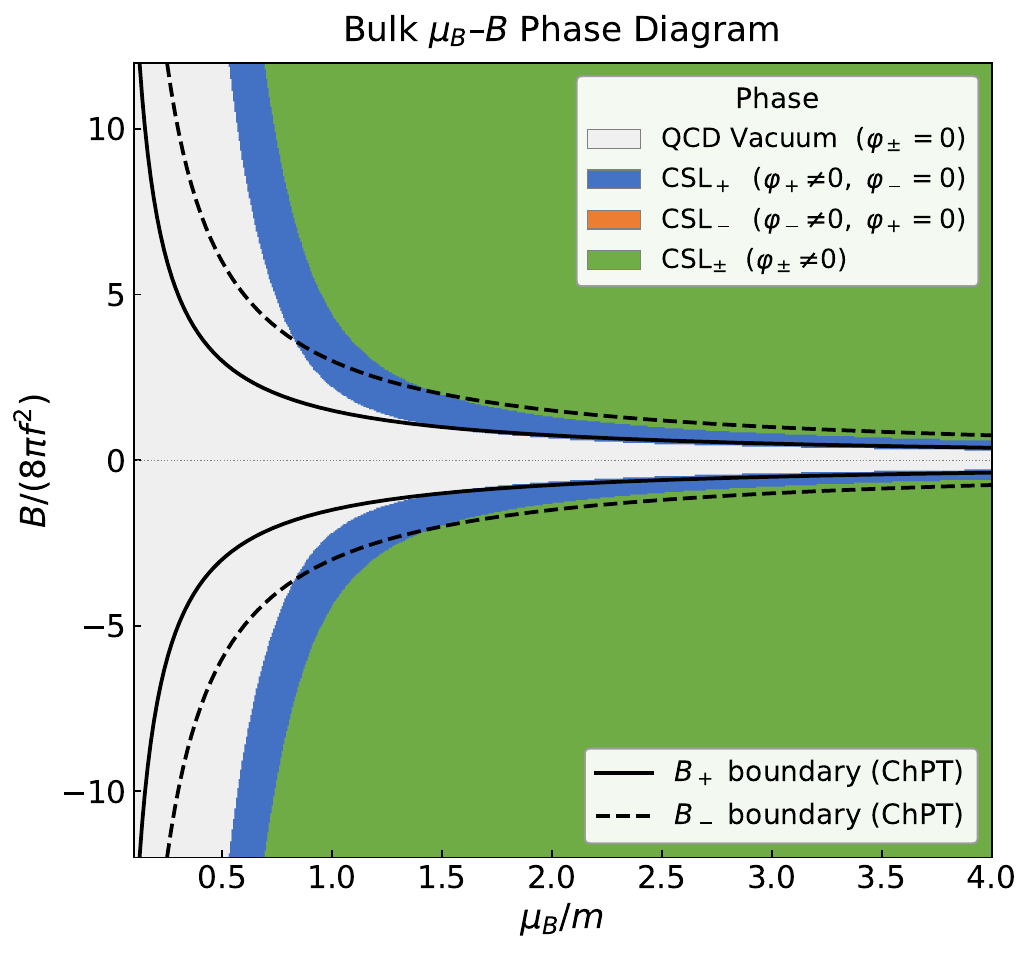}%
  \includegraphics[width=0.33\textwidth]{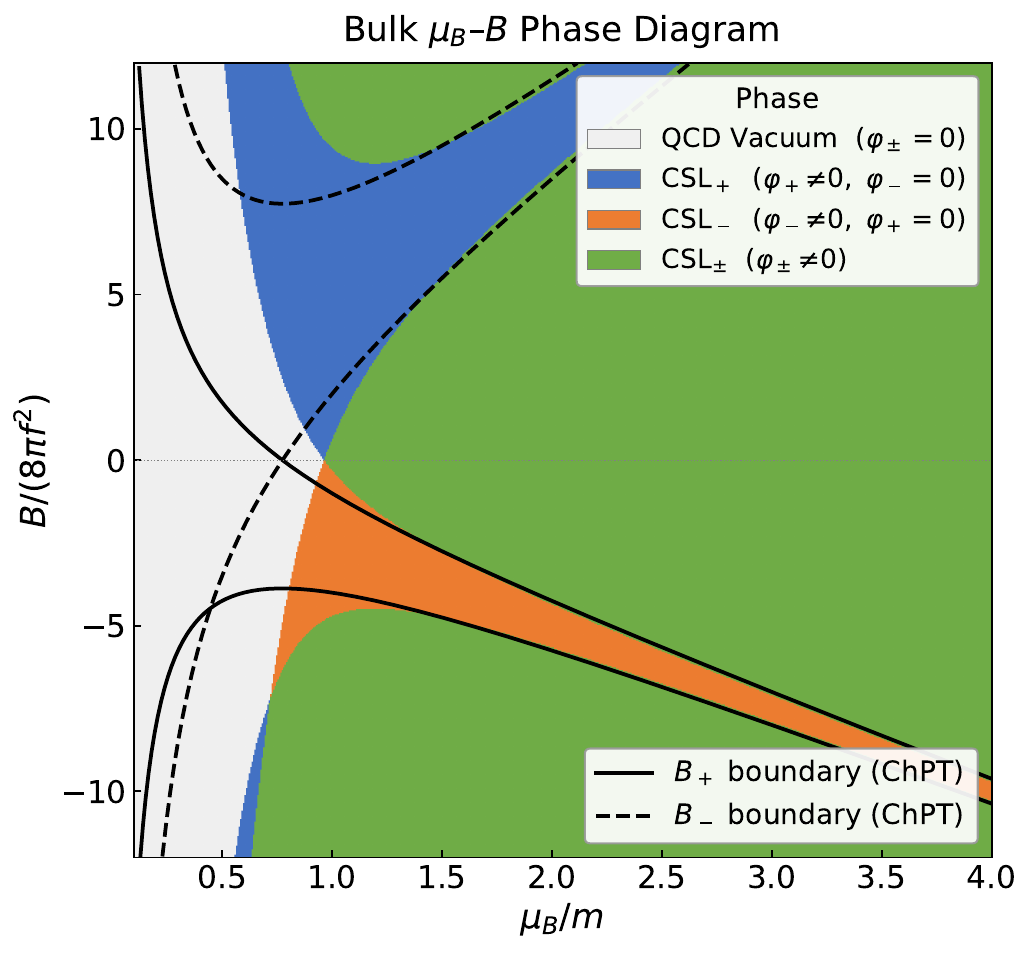}%
  \includegraphics[width=0.33\textwidth]{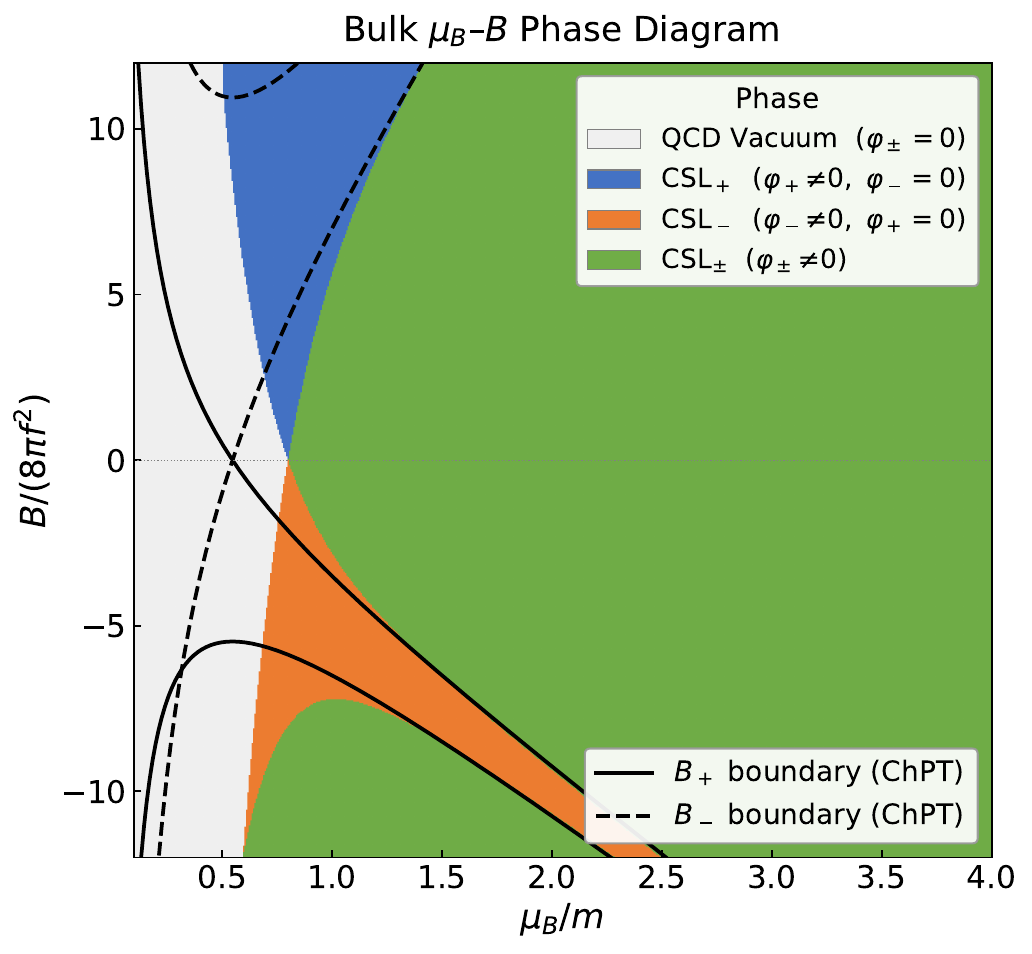}
  \caption{Bulk phase diagrams in the $\mu_B$--$B$ plane, with axes shown
    as the dimensionless ratios $\mu_B/m$ and $B/(8\pi f^2)$.
    Left: $\Omega/(N_c\,8\pi f^2) = 0$, where no $\mathrm{CSL}_-$ band appears since $|B_+| \geq |B_-|$.
    Middle: $\Omega/(N_c\,8\pi f^2) = 2.5$ and Right: $\Omega/(N_c\,8\pi f^2) = 5.0$, where
    rotation breaks this inequality and a $\mathrm{CSL}_-$ band emerges at negative~$B$,
    widening as $\Omega$ increases.
    Solid (dashed) curves are the ChPT $B_+$ ($B_-$) boundaries, given by
    $(\mu_B/m)\,|B_\pm|/(8\pi f^2) = 1$.}
  \label{fig:bulk_phase_mu_B}
\end{figure*}

\begin{figure*}[t]
  \centering
  \includegraphics[width=0.48\textwidth]{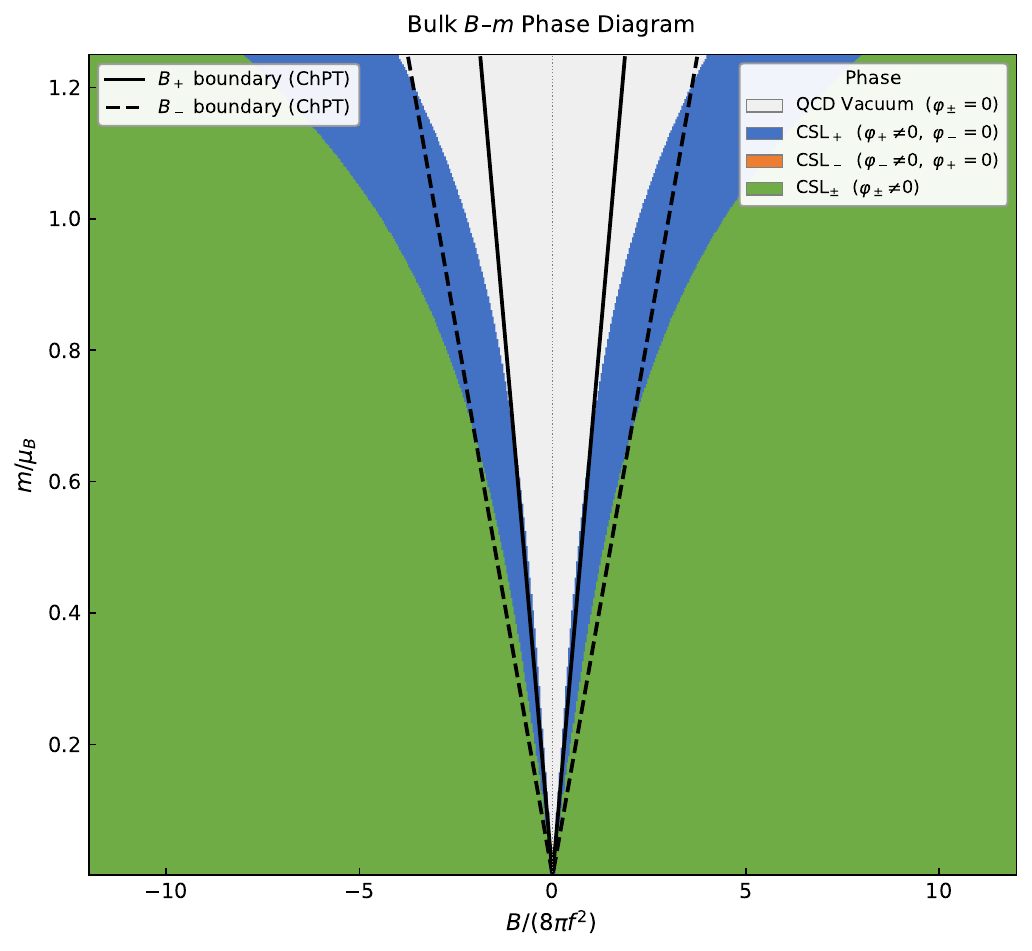}
  \includegraphics[width=0.48\textwidth]{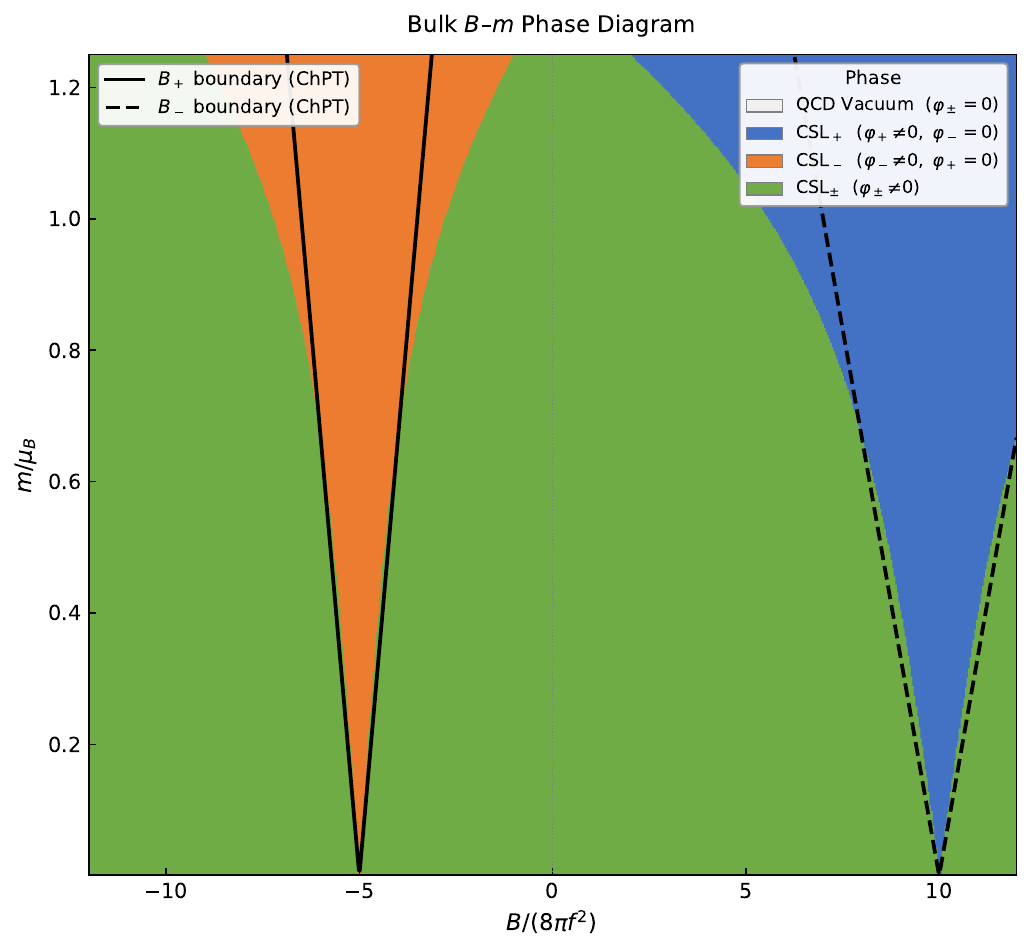}%
  \caption{Bulk phase diagrams in the $m$--$B$ plane at fixed $\mu_B = 1.5$,
    with axes shown as the dimensionless ratios $B/(8\pi f^2)$ and $m/\mu_B$.
    Left: $\Omega/(N_c\,8\pi f^2) = 0$.  Right: $\Omega/(N_c\,8\pi f^2) = 5.0$.
    As $m$ grows the CSL regions shrink and the vacuum expands, consistent
    with the condition $|B_\pm|/(8\pi f^2) \geq m/\mu_B$.
    Rotation ($\Omega \neq 0$) shifts and distorts the boundaries,
    opening a $\mathrm{CSL}_-$ window at negative~$B$.
    Solid (dashed) curves are the ChPT $B_+$ ($B_-$) boundaries, given by
    $|B_\pm|/(8\pi f^2) = m/\mu_B$.}
  \label{fig:bulk_phase_mpi_B}
\end{figure*}

The bulk phase structure is determined numerically by evaluating $E_\mathrm{min}(|\mathcal{B}_\pm|, \mu_\pm)$ from the on-shell Hamiltonian\footnote{Note that $\mu_\pm = \mu_B/N_c$.} \eqref{eq:bulk_smallm_energy} and checking the sign for each of the $\pm$ sectors independently.
Phase diagrams obtained from this numerical procedure are shown in Figures~\ref{fig:bulk_phase_mu_Omega}--\ref{fig:bulk_phase_mpi_B}.
The four phases (QCD vacuum, $\mathrm{CSL}_+$, $\mathrm{CSL}_-$, and $\mathrm{CSL}_\pm$) are identical to those of the ChPT analysis (see Table~\ref{tab:chpt_csl_phases}), but the boundaries are shifted by the bulk backreaction encoded in $\tilde{f}_{ij}$.
The ChPT analytic boundaries $\mu_B|B_\pm| = 8\pi f^2 m$ are overlaid as solid ($B_+$) and dashed ($B_-$) curves for direct comparison.

\begin{figure}[t]
  \centering
  \includegraphics[width=0.48\textwidth]{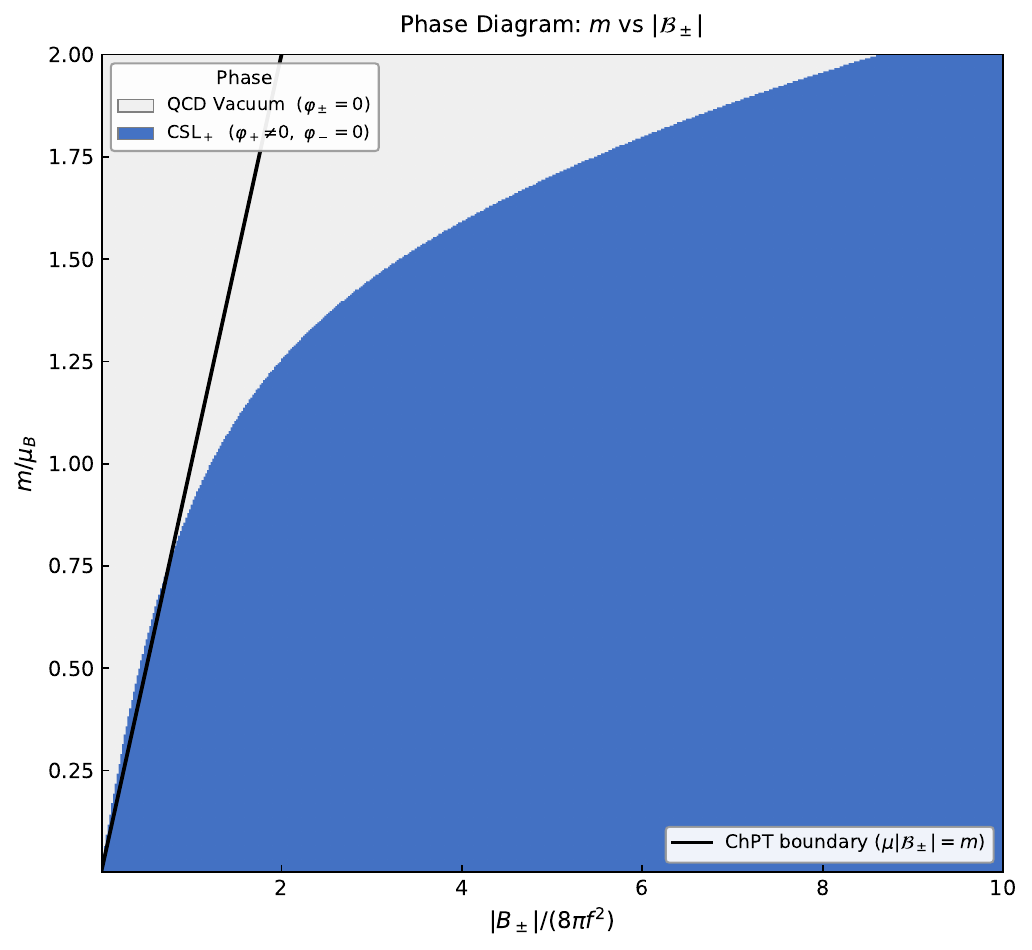}
  \includegraphics[width=0.48\textwidth]{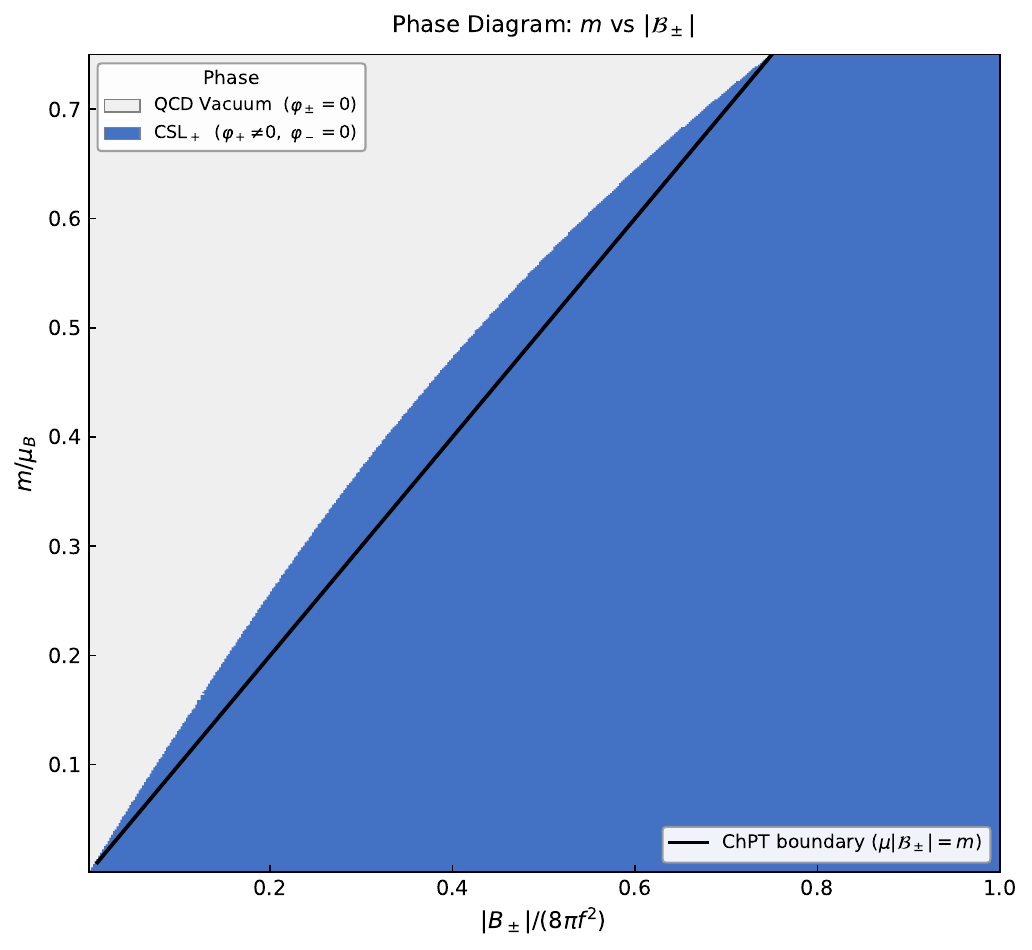}
  \caption{Phase diagrams in the $|\mathcal{B}_\pm|/(8\pi f^2)$--$m/\mu_B$ plane at fixed
    baryon chemical potential $\mu_B$.
    The coloured region indicates the CSL phase ($E_\mathrm{min} < 0$);
    the uncoloured region is the QCD vacuum.
    The solid curve is the ChPT phase boundary
    $\mu_B\,|B_\pm| = 8\pi f^2\,m$,
    or equivalently $m/\mu_B = |B_\pm|/(8\pi f^2)$,
    above which the CSL phase is no longer energetically favoured.
    The left diagram covers a larger range over the $|B_\pm|/(8\pi f^2)$--$m/\mu_B$ plane, versus the right diagram, which is zoomed in on the bottom left corner.
    The right diagram shows that accounting for all the boundary harmonics makes the formation of CSL phases easier for weak magnetic fields (vorticities) and small pion masses.
    }
  \label{fig:phase_mpi_vs_bpm}
\end{figure}

For small $m$ and $B_\pm$, accounting for higher-order modes in the $x^3$ direction demonstrates that CSL phases form more readily.
This is visible in the enlarged region of Figure \ref{fig:phase_mpi_vs_bpm}, as well as in Figure \ref{fig:bulk_phase_mpi_B} near $m \approx 0$ and $B_\pm \approx 0$. 
In environments with competing rotational and magnetic fields, offsetting one field against the other enhances the CSL phases under strong magnetic fields or rotation.
This occurs because even in the presence of a strong magnetic field, a counter-rotation can compensate for it, driving either $B_+ \approx 0$ or $B_- \approx 0$. 
Nevertheless, this result should be interpreted with caution, as the effect is relatively small.
Furthermore, because we are ostensibly working in the large-$N_c$ limit, it remains unclear whether this enhancement stems from the underlying QCD theory or is an artifact of the large-$N_c$ approximation.

\subsection{Large/Small pion mass phase approximations}

\paragraph{Small pion mass}
For small $m$ and/or relatively large $B$, we assume we can neglect higher order terms of $\phi$.
Concretely, this approximation is an approximation of the amplitude function, $\am(x, k) = x + \mathcal O(x^2)$.
The bulk equations of motion turn to be equivalent to the massless case.
This leaves us with an additional mass term.
We can use \eqref{eq:phi_pm_amplitude} to relate $\pd_3\phi$ to $k$ as $\pd_3\phi = \pm 2m/k$:
\begin{equation}
    \label{eq:small_mpi_H_mass}
    \begin{aligned}
        H_\mathrm{mass} &= \sum_\pm\frac{f^2 m^2}{2} \int d^3x (1 - \cos(\phi_\pm)) =  V_3 \sum_\pm\frac{f^2 m^2}{2}. \\
    \end{aligned}
\end{equation}
Effectively, we obtain an on-shell Hamiltonian that is very similar to the $m = 0$ case plus \eqref{eq:small_mpi_H_mass}.
Note that for vacuum ($\phi_\pm = 0$), the mass term disappears for each $\phi_\pm$ sector.
We write out the total condensation-phase, on-shell Hamiltonian as
\begin{equation}
    \begin{aligned}
        \frac{H_{4D}}{V_3} &=  \sum_{\pm}\left( \frac 14 \pd_i\phi_\pm\tilde f^2_{ij\pm}\pd_j\phi_\pm - \frac\pi4 f^2 \mu_\pm \mathcal B_{i\pm} \pd_i\phi_\pm + \frac 12 f^2 m^2\right).\\
    \end{aligned}
\end{equation}
So the minimum energy (density) CSL state is approximately.
\begin{equation}
    \label{eq:bulk_smallm_energy}
    \sum_\pm \left(
    -\frac{3\pi f^2}{8} \frac{ \mu_\pm^2|\mathcal B_\pm| (\cosh (\pi  |\mathcal B_\pm|)-1)}{4 \sinh
    (\pi  |\mathcal B_\pm|)-\pi  |\mathcal B_\pm|}+ \frac 12 f^2 m^2
    \right).
\end{equation}
This is only preferable to the QCD vacuum if $H_{4D}/V_3 < 0$, so we have two independent conditions for a maximum of $4$ possible phases similar to Section \ref{sec:chpt_csl_nf2}.
For weak magnetic fields and small pion mass, the result reduces to a more accurate result than that found in \cite{Amano:2025iwi}.
Where one simply makes a substitution of $f\rightarrow \tilde f$, indicating that a Witten-Sakai-Sugimoto-CSL is more difficult to create with respect to a ChPT-CSL.
We can extend this approximation to the large magnetic field case in which $\tilde f$ become linear in $|\mathcal B|$.
Explicitly, setting \eqref{eq:bulk_smallm_energy} to zero sector by sector fixes the critical field $|\mathcal B_\pm|$ above which the CSL$_\pm$ is preferred over the QCD vacuum.
Expanding \eqref{eq:bulk_smallm_energy} to $\mathcal O(\mathcal B_\pm^2)$, appropriate for $|\mathcal B_\pm| \ll 1$, gives an energy density $-\frac{\pi^2}{16}f^2\mu_\pm^2\mathcal B_\pm^2 + \frac12 f^2m^2$ and hence
\begin{equation}
    \label{eq:bulk_smallm_smallB_Bc}
    |\mathcal B_\pm| \gtrsim \frac{2\sqrt 2\, m}{\pi \mu_\pm}
    \qquad \Longleftrightarrow \qquad
    \mu_B |B_\pm| \gtrsim 2\sqrt 2\, \pi^2 f^2 m
    = \frac{\pi}{2\sqrt 2}\left( 8\pi f^2 m \right).
\end{equation}
The approximation of the original $\am$ function can be seen to change the CSL bounds by around $11\%$.
We can make a better approximation by multiplying the mass term energy density term by $8/\pi^2$, restoring the equality between WSS and ChPT for small magnetic fields.
In the opposite regime, $|\mathcal B_\pm| \gg 1$, the ratio in \eqref{eq:bulk_smallm_energy} saturates, so that $\tilde f^2_{ij\pm}$ is linear in $|\mathcal B_\pm|$.
Balancing against the mass term then yields
\begin{equation}
    \label{eq:bulk_smallm_largeB_Bc}
    |\mathcal B_\pm| \gtrsim \frac{16\, m^2}{3\pi \mu_\pm^2}
    \qquad \Longleftrightarrow \qquad
    \mu_B^2 |B_\pm| \gtrsim \frac{16 \pi^2 N_c f^2 m^2}{3},
\end{equation}
where we have used $\mu_\pm = \mu_B/N_c$.
The two regimes scale differently: the weak-field threshold is linear in $m$ and falls as $\mu_B^{-1}$, whereas the strong-field threshold is quadratic in $m$ and falls as $\mu_B^{-2}$, with an explicit $N_c$ enhancement inherited from the Chern--Simons coupling.
Since each sector carries its own $\mu_\pm$ and $\mathcal B_\pm$, \eqref{eq:bulk_smallm_smallB_Bc} and \eqref{eq:bulk_smallm_largeB_Bc} are two independent conditions, reproducing the four-phase structure of Table~\ref{tab:chpt_csl_phases} with boundaries shifted relative to ChPT.
Notice that for large magnetic fields, it scales like $m^2$, which is harsher than the $m^1$ behavior of ChPT.

\paragraph{Large pion mass}

For large $\mu_B$ or $m$, numerically we found that the energy minimizing $k$ value was $k \approx 1$.
Using the full analytic profile for $\phi$ would require a different numerical approach.
Instead, we use a crude approximation of
$\am(x,1)=\mathrm{gd}\,x\approx2\arctan\sinh(mx)$.
Let $\hat n$ be an arbitrary unit vector defining the direction of the CSL modulation and define
$s=\hat n\cdot\mathbf{x}$:
\begin{equation}
    2\arctan\sinh(ms)
    \approx
    \begin{cases}
        -\pi, & s<-\dfrac{\pi}{2m}\\
        2ms, & |s|\le\dfrac{\pi}{2m}\\
        \pi, & s>\dfrac{\pi}{2m}.\\
    \end{cases}
\end{equation}
This approximation gives an incorrect ground-state energy, but its linearity allows the bulk solutions to be evaluated analytically:

\begin{equation}
    \label{eq:large_mpi_H_mass}
    \frac{H_\mathrm{mass}}{V_\perp}
    =
    \sum_\pm
    \frac{\pi}{2}f^2m.
\end{equation}
Here the cosine term in the pion mass action evaluates to zero.

Analogously to \eqref{eq:bulk_chiral_4D_hamiltonian}, the Hamiltonian becomes
\begin{equation}
    \begin{aligned}
        H_{4D} &= V_\perp \sum_\pm \Biggl( \frac{\pi}{4m} \hat n_{i\pm} \tilde f^2_{ij\pm} \hat n_{j\pm} (\hat n_{k\pm}\partial_k\phi_\pm)^2 - \frac{\pi^2}{4m} f^2 \mu_\pm \mathcal B_{i\pm} \hat n_{i\pm} (\hat n_{k\pm}\partial_k\phi_\pm) + \frac{\pi}{2}f^2m \Biggr) \\
        &= V_\perp \sum_\pm \left( \pi m\, \hat n_{i\pm}\tilde f^2_{ij\pm}\hat n_{j\pm} - \frac{\pi^2}{2} f^2 \mu_\pm \left|\mathcal B_{i\pm}\hat n_{i\pm}\right| + \frac{\pi}{2}f^2m \right).
    \end{aligned}
\end{equation}
The CSL phase is stable for $H_{4D}<0$ for either $\pm$ summand, just like sec. \ref{sec:chpt_csl_nf2}.
Note that we normalize with respect to $V_\perp$, the spatial volume transverse to the modulation direction $\hat{n}_\pm$, rather than $L_\parallel = \pi/m$, because the energy density is compact along $\hat{n}_\pm$ over the interval $\left[-\frac{\pi}{2m},\,\frac{\pi}{2m}\right]$.
Nevertheless, since we do not have any free variables to vary to minimize the energy, normalizing or not does not change the stabilization condition:
\begin{equation}
    m\,
    \hat n_{i\pm}
    \tilde f^2_{ij\pm}
    \hat n_{j\pm}
    -
    \frac{\pi}{2}
    f^2
    \mu_\pm
    \left|\mathcal B_{i\pm}\hat n_{i\pm}\right|
    +
    \frac12
    f^2m
    <0.
\end{equation}

When the modulation is parallel to the magnetic field,
$\hat n=\hat{\mathcal B}_\pm$, this reduces to
\begin{equation}
    m\frac{2\pi}{3}|\mathcal B_\pm|
    -
    \frac{\pi}{2}
    \mu_\pm
    |\mathcal B_\pm|
    +
    \frac12m
    <0\,.
\end{equation}
This implies that
\begin{equation}
    |\mathcal B_\pm|
    \gtrsim
    \frac{3m}{\pi(3\mu_\pm-4m)},
\end{equation}
if $\mu_\pm$ is approximately greater than $4m/3$; otherwise, the corresponding CSL$_\pm$ phase does not exist.

Beyond the low-energy regime, we also numerically find support for a maximum pion mass at which the CSL phase is stable.
One can see in the following plot that around some $\mathcal B_\pm$ value the phase boundary becomes horizontal up to reasonable numerical accuracy.

It can be seen in Figures \ref{fig:bulk_phase_mu_Omega} and \ref{fig:bulk_phase_mu_B} that the magnitude of the chemical potential is significant even when it is small.
In the regime around the pion-mass scale, the upper limit of the chemical potential cannot be neglected.
Consequently, the phase boundaries between WSS and ChPT differ significantly.
On the other hand, for large chemical potentials the anomalous term dominates; since both WSS and ChPT share the same anomalous term, their corresponding ground states become approximately equivalent.

\section{Conclusion and discussions}
\label{sec:conclusion}
In this work, we have studied the CSL in holographic QCD matter in the presence of a background magnetic field and rotation.
To leading order in the local velocity, the rotational effect is encoded by an effective magnetic component of the background $\mathrm{U}(1)_B$ gauge field 
proportional to $\mu_B \vec{\Omega}$.
This allows us to analyze rotating dense QCD matter within the framework of the Sakai-Sugimoto model.

For the $N_f=1$ case, we have shown that the $\eta$ meson in the rotating matter obeys the sine-Gordon equation and that the corresponding CSL becomes energetically favored above a critical value of a combination of $\mu_B$ and $\vec{\Omega}$.
Then the rotating CSL becomes a nontrivial ground state in the gravity dual of large-$N_c$ QCD.
We have also provided a brane interpretation of this rotational CSL. 
Unlike the purely magnetic case studied in our previous paper \cite{Amano:2025iwi}, where the corresponding objects provide charge only for the (partially) dissolved D4-branes accompanied with the non-zero instanton density, the rotational setup naturally involves charges for dissolved D6-branes sourced by the background $\mathrm{U}(1)_B$ field and kink.
Thus, rotation gives a distinct geometric realization of the CSL in the holographic setup.
For the $N_f=2$ case, we have included both the magnetic field in $\mathrm{U}(1)_{\mathrm{em}}$ 
and rotation and focused on the neutral pion and eta sectors.
In terms of the combinations $\phi_\pm = \frac{1}{f} (\eta \pm \pi^0)$, 
the system reduces to two decoupled sine-Gordon sectors with the topological term given by linear combinations of $\vec{B}$ and $\mu_B \vec{\Omega}$. 
This leads to a richer phase structure than in the previous analysis, which considered only the effects of $\vec{B}$.
Depending on the values of the magnetic field, angular velocity, and baryon chemical potential, we find phases in which the CSL forms in the $\phi_+$ sector, the $\phi_-$ sector, both sectors, or neither sector (the trivial vacuum).

At the holographic bulk level, we solved the gauge field equations in the presence of general magnetic backgrounds and derived the corresponding on-shell effective Hamiltonian. 
A notable result is that the meson decay constant is replaced by an anisotropic, field dependent matrix $\tilde{f}_{ij}$.
This is a generalization of our previous result where the bulk analysis in the $\mathrm{U}(1)_{\mathrm{em}}$ magnetic background $B$ led to a $B$-dependent effective pion decay constant.
This characteristic is naturally captured through our consistent five-dimensional bulk analysis.

In this work, the effects of rotation have been incorporated effectively through a
background $\mathrm{U}(1)_B$ gauge field that encodes the vorticity. 
We stress that this background field prescription is technically much more tractable than constructing
and analyzing the holographic system directly in rotating spacetime geometries \cite{Chen:2020ath, Braga:2022yfe}.
A fully geometric treatment of rotation would provide an important extension of the present analysis.

There remain several important directions for future work. 
First, it would be interesting to carry out a fully systematic numerical analysis of the massive bulk equations in the regime where the magnetic field, rotation, and meson mass are all comparable. 
Second, the present analysis has focused on neutral mesons only. 
Including charged pions may reveal additional phase structure.
It is also important to clarify how the CSL phase is connected to the domain-wall Skyrmion phase in the presence of both magnetic field and rotation. 
Finally, it would be interesting to study finite temperature effects and possible implications for dense QCD matter in neutron stars and related strongly magnetized, rotating environments.
We will return to these issues in future studies.

\begin{acknowledgments}
This work is supported in part by Japan Society for the Promotion of
 Science (JSPS) KAKENHI [Grants No. JP22H01221 and JP23K22492 (ME and
 MN), JP25K07324 (SS)] and the WPI program ``Sustainability with Knotted
 Chiral Meta Matter (WPI-SKCM$^2$)'' at Hiroshima University (ME and
 MN). 
 MA was provided funding by the National Institute of Technology, Oyama College.
\end{acknowledgments}

\appendix

\section{Reduction to ChPT in the massive case} \label{app:Reduction_to_ChPT}

We start with the bulk equations of motion without the CS term.
\begin{equation}
  \begin{aligned}
      \pd_3 (u^3 F_{z3}) &= - m^2 \sum_\pm  \sin(\phi_\pm) \tau^\pm, \\
      u^{-1} \pd_3 F_{3t} +  \pd_z \left(u^3 F_{zt}\right) &= 0, \\
      \pd_z \left(u^3 F_{z3}\right) &= 0,\\
  \end{aligned}
\end{equation}
Note, that we use the $\tau^\pm$ basis implicitly.
The last equation and boundary conditions imply that 
\begin{equation}
    A_3 = -\frac 1\pi \pd_3 \phi \xi.
\end{equation}
To be consistent with the first equation, we find that
\begin{equation}
    \pd_3^2 \phi = m^2 \sin(\phi).
\end{equation}
Implying that $\phi$ admits the solution \eqref{eq:phi_pm_amplitude}.
The second equation and $A_t$ boundary conditions are satisfied for $A_t = \mu$.
This allows us to evaluate the action explicitly.
\begin{equation}
  \begin{aligned}
      H_\mathrm{4D} &= f^2 V_2 \int dx^3 \tr \left(
      \frac 14 (\pd_3\phi)^2 -
      \frac\pi{12} \mu \mathcal B_i \pd_i\phi
      + \frac {m^2}4 (2
      - U - U^\dagger) \right).
  \end{aligned}
\end{equation}
Evaluating the integrals and averaging out the energy, we have
\begin{equation}
    \frac{1}{f^2}\frac{H_\mathrm{4D}}{T V_2} =: \frac{1}{f^2}\sigma = \frac{m \left(4 \pi  m \left(\left(k^2-1\right) \mathcal K(k_\pm)+2
    E(k_\pm)\right)-B_\pm k_\pm \mu_\pm \right)}{4 \pi  k_\pm^2 \mathcal K(k_\pm)}
\end{equation}
where we subbed in for $B$ and $N_c=3$ matching section \ref{sec:chpt_csl_nf2}.

\section{Bulk solutions for the massless case}
\label{app:massless_bulk_solutions}

In this appendix we solve the coupled ODEs \eqref{eq:massless_coupled_ODEs} for the bulk fields in the massless case $m=0$, subject to the boundary conditions \eqref{eq:boundary_conditions}.

\paragraph{Change of coordinates}
It is convenient to pass to the $\xi = \arctan z \in (-\pi/2,\pi/2)$ coordinate, in which $K(z)\,\pd_z = \pd_\xi$.
The coupled ODEs \eqref{eq:massless_coupled_ODEs} then become
\begin{equation}
    \begin{aligned}
        \pd_\xi F_{\xi t} &= -\mathcal B_i \delta^{ij} F_{\xi j},\\
        \pd_\xi F_{\xi i} &= -\mathcal B_i F_{\xi t}.\\
    \end{aligned}
\end{equation}
Eliminating $F_{\xi i}$ yields a decoupled equation for $F_{\xi t}$:
\begin{equation}
    \pd_\xi^2 F_{\xi t} - |\mathcal B|^2 F_{\xi t} = 0,
\end{equation}
where $|\mathcal B|^2 = \mathcal B_i \delta^{ij} \mathcal B_j$.

\paragraph{Decoupling in the $\tau^\pm$ basis}
Since we work in the Cartan subalgebra of $\mathfrak{u}(2)$, we introduce the projector basis $\tau^\pm := \frac 12 \left(\tau^0 \pm \tau^3\right)$.%
\footnote{In the $\tau^\pm$ basis the equations of motion decouple because $(\tau^+)^2 = \tau^+$, $(\tau^-)^2 = \tau^-$, and $\tau^+\tau^- = 0$. Hence any analytic function satisfies $f(a_b \tau^b) = f(a_+)\tau^+ + f(a_-)\tau^-$ with $b\in\{+,-\}$.}
Setting $\mathcal{B}_{i\pm} = \mathcal{B}_{i0} \pm \mathcal{B}_{i3}$, each $\pm$ sector satisfies
\begin{equation}
    \pd_\xi^2 F_{\xi t\pm} - |\mathcal B_\pm|^2 F_{\xi t\pm} = 0.
\end{equation}
The boundary conditions \eqref{eq:boundary_conditions} require $A_t(\xi=\pm\pi/2) = \mu_\pm$, which is even in $\xi$.
Since $F_{\xi t} = \pd_\xi A_t$, this means $F_{\xi t}$ must be odd in $\xi$, so we discard the even (cosh) solution and retain only
\begin{equation}
    F_{\xi t \pm} = \zeta_\pm \sinh (|\mathcal B_\pm|\xi),
\end{equation}
where $\zeta_\pm$ are constants of integration to be fixed by the boundary conditions.

\paragraph{Integration to obtain $A_t$ and $F_{\xi i}$}
Integrating $F_{\xi t\pm} = \pd_\xi A_{t\pm}$ gives
\begin{equation}
    A_{t\pm} = \frac{\zeta_\pm}{|\mathcal B_\pm|} 
    \left(\cosh (|\mathcal B_\pm|\xi) - \cosh (|\mathcal B_\pm|\pi/2)\right) + \mu_\pm,
\end{equation}
where we have imposed $A_{t\pm}(\xi=\pm\pi/2) = \mu_\pm$.
From $\pd_\xi F_{\xi i\pm} = -\mathcal{B}_{i\pm} F_{\xi t\pm}$,
\begin{equation}
    F_{\xi i\pm} = -\zeta_\pm \frac{\mathcal B_{i\pm}}{|\mathcal B_\pm|} \cosh (|\mathcal B_\pm|\xi) + C_{i\pm},
\end{equation}
where $C_{i\pm}$ are integration constants. Integrating once more yields $A_{i\pm}$:
\begin{equation}
    A_{i\pm} = -\zeta_\pm \frac{\mathcal B_{i\pm}}{|\mathcal B_\pm|^2} \sinh (|\mathcal B_\pm|\xi) + C_{i\pm}\xi.
\end{equation}

\paragraph{Fixing integration constants from boundary conditions}
The spatial boundary condition $A_{i\pm}(\xi=\pm\pi/2) = -\frac{1}{2}\pd_i\phi_\pm$ gives
\begin{equation}
    C_{i\pm} = \frac{2}{\pi}\left( \zeta_\pm \frac{\mathcal B_{i\pm}}{|\mathcal B_\pm|^2} \sinh (|\mathcal B_\pm|\pi/2) - \frac{1}{2}\pd_i\phi_\pm \right),
\end{equation}
where $\phi_\pm = \phi_0 \pm \phi_3$.
The equation of motion for $F_{\xi t}$ also requires $C_{i\pm}\mathcal{B}_{i\pm} = 0$, which fixes
\begin{equation}
    \begin{aligned}
        \zeta_\pm &= \frac{\frac{1}{2}\mathcal B_{i\pm}\pd_i\phi_\pm} {\sinh(|\mathcal B_\pm|\pi/2)},\\
        C_{i\pm}  &= \frac{2}{\pi}\left( \frac{\mathcal B_{i\pm}}{|\mathcal B_\pm|^2} \mathcal B_{j\pm}\frac{1}{2}\pd_j\phi_\pm - \frac{1}{2}\pd_i\phi_\pm \right).
    \end{aligned}
\end{equation}

\paragraph{Compact form via the transverse projector}
Introducing the projector onto the plane perpendicular to $\mathcal{B}_\pm$,
\begin{equation}
    \mathcal{P}_{ij\pm} := \delta_{ij} - \hat{\mathcal{B}}_{i\pm}\hat{\mathcal{B}}_{j\pm},
\end{equation}
where $\hat{\mathcal{B}}_\pm = \mathcal{B}_\pm/|\mathcal{B}_\pm|$, the integration constant simplifies to $C_{i\pm} = -\frac{1}{\pi}\mathcal{P}_{ij\pm}\pd_j\phi_\pm$.
The complete on-shell bulk fields are then
\begin{equation}
    \label{eq:app_massless_bulk_solution}
    \begin{aligned}
        F_{\xi i\pm} &= -\left( \frac{1}{2}|\mathcal B_\pm|\frac{\cosh(|\mathcal{B}_\pm|\xi)}{\sinh(|\mathcal{B}_\pm|\pi/2)} \hat{\mathcal B }_{i\pm} \hat{\mathcal B}_{j\pm} + \frac{1}{\pi} \mathcal{P}_{ij\pm} \right)  \pd_j\phi_\pm, \\
        A_{i\pm} &= - \left(\frac{1}{2}\frac{\sinh(|\mathcal{B}_\pm|\xi)}{\sinh(|\mathcal{B}_\pm|\pi/2)} \hat{\mathcal B }_{i\pm} \hat{\mathcal B}_{j\pm} + \frac{1}{\pi} \mathcal{P}_{ij\pm} \xi \right)  \pd_j\phi_\pm, \\
        F_{\xi t\pm} &= \frac{1}{2}|\mathcal B_\pm|\frac{\sinh(|\mathcal{B}_\pm|\xi)}{\sinh(|\mathcal{B}_\pm|\pi/2)} \hat{\mathcal{B}}_{j\pm}\pd_j\phi_\pm,  \\
        A_{t\pm} &= \frac{1}{2}\frac{\cosh(|\mathcal{B}_\pm|\xi) - \cosh(|\mathcal{B}_\pm|\pi/2)}{\sinh(|\mathcal{B}_\pm|\pi/2)} \hat{\mathcal{B}}_{j\pm}\pd_j\phi_\pm + \mu_\pm.
    \end{aligned}
\end{equation}
These are the on-shell bulk solutions stated in eq.~\eqref{eq:massless_bulk_solution} of the main text.

\bibliographystyle{jhep}
\bibliography{reference}

\end{document}